%% file: icml2020_conference.tex
\documentclass{article}

\usepackage{microtype}
\usepackage{graphicx}
\usepackage{subfigure}
\usepackage{booktabs} 

\usepackage{color}
\usepackage{mathtools}
\input{math_commands.tex}

\usepackage{subfigure}
\usepackage{multirow}
\usepackage{makecell}
\usepackage{array, booktabs, arydshln, xcolor}
\usepackage{amssymb}
\usepackage{graphbox}
\usepackage{url}

\usepackage{amsthm}

\usepackage{thmtools}
\usepackage{thm-restate}
\usepackage{hyperref}
\usepackage{cleveref}

\allowdisplaybreaks

\usepackage{color}
\usepackage{mathtools}

\usepackage{subfigure}
\usepackage{multirow}
\usepackage{makecell}
\usepackage{array, booktabs, arydshln, xcolor}
\usepackage{amssymb}
\usepackage{graphbox}
\usepackage{caption}
\usepackage{wrapfig}
\allowdisplaybreaks

\newcommand{\shortn}{\textup{\texttt{-}}}
\newcommand{\shorte}{\textup{\texttt{=}}}

\newcommand{\name}{ROMA}
\newcommand{\Tau}{\mathrm{T}}

\usepackage{hyperref}

\usepackage[accepted]{icml2020}

\icmltitlerunning{ROMA: Role-Oriented Multi-Agent Reinforcement Learning}

\begin{document}

\twocolumn[
\icmltitle{ROMA: Multi-Agent Reinforcement Learning with Emergent Roles}



\icmlsetsymbol{equal}{*}

\begin{icmlauthorlist}
\icmlauthor{Tonghan Wang}{thu}
\icmlauthor{Heng Dong}{thu}
\icmlauthor{Victor Lesser}{umass}
\icmlauthor{Chongjie Zhang}{thu}
\end{icmlauthorlist}

\icmlaffiliation{thu}{IIIS, Tsinghua University, Beijing, China}
\icmlaffiliation{umass}{University of Massachusetts, Amherst, USA}

\icmlcorrespondingauthor{Tonghan Wang}{tonghanwang1996@gmail.com}

\icmlkeywords{Multi-Agent Reinforcement Learning, Role, Division of Labor, Sub-Task Specialization}

\vskip 0.3in
]



\printAffiliationsAndNotice{}  

\begin{abstract}
The role concept provides a useful tool to design and understand complex multi-agent systems, which allows agents with a similar role to share similar behaviors. However, existing role-based methods use prior domain knowledge and predefine role structures and behaviors. In contrast, multi-agent reinforcement learning (MARL) provides flexibility and adaptability, but less efficiency in complex tasks. In this paper, we synergize these two paradigms and propose a role-oriented MARL framework (\name). In this framework, roles are emergent, and agents with similar roles tend to share their learning and to be specialized on certain sub-tasks. To this end, we construct a stochastic role embedding space by introducing two novel regularizers and conditioning individual policies on roles. Experiments show that our method can learn specialized, dynamic, and identifiable roles, which help our method push forward the state of the art on the StarCraft II micromanagement benchmark. Demonstrative videos are available at \url{https://sites.google.com/view/romarl/}.
\end{abstract}

\input{1-Introduction.tex}
\input{2-Preliminaries.tex}
\input{3-Methods.tex}
\input{4-RelatedWorks.tex}
\input{5-Experiments.tex}
\input{6-ClosingRemarks.tex}
\section*{Acknowledgements}

We gratefully acknowledge Jin Zhang for his valuable discussions. We also would like to thank the reviewers for their detailed and constructive feedback. This work is partially supported by the sponsorship of Guoqiang Institute at Tsinghua University.


\bibliography{icml2020.bib}
\bibliographystyle{icml2020}

\appendix
\onecolumn
\input{A1-Math.tex}
\input{A2-Architecture.tex}
\input{A3-Experiments.tex}
\input{A4-RelatedWorks.tex}





\end{document}

%% file: math_commands.tex
\usepackage{amsmath,amsfonts,bm}

\def\eqref#1{equation~\ref{#1}}

\def\1{\bm{1}}

\def\va{{\bm{a}}}

\DeclareMathAlphabet{\mathsfit}{\encodingdefault}{\sfdefault}{m}{sl}
\SetMathAlphabet{\mathsfit}{bold}{\encodingdefault}{\sfdefault}{bx}{n}

\newcommand{\KL}{D_{\mathrm{KL}}}



%% file: 1-Introduction.tex
\section{Introduction}

Many real-world systems can be modeled as multi-agent systems (MAS), such as autonomous vehicle teams~\cite{cao2012overview}, intelligent warehouse systems~\cite{nowe2012game}, and sensor networks~\cite{zhang2011coordinated}. Cooperative multi-agent reinforcement learning (MARL) provides a promising approach to developing these systems, allowing agents to deal with uncertainty and adapt to the dynamics of an environment. In recent years, cooperative MARL has achieved prominent progress, and many deep methods have been proposed~\cite{foerster2018counterfactual, sunehag2018value, rashid2018qmix, son2019qtran, vinyals2019grandmaster, wang2020learning, baker2020emergent}. 
\begin{figure}
\includegraphics[height=0.35\linewidth]{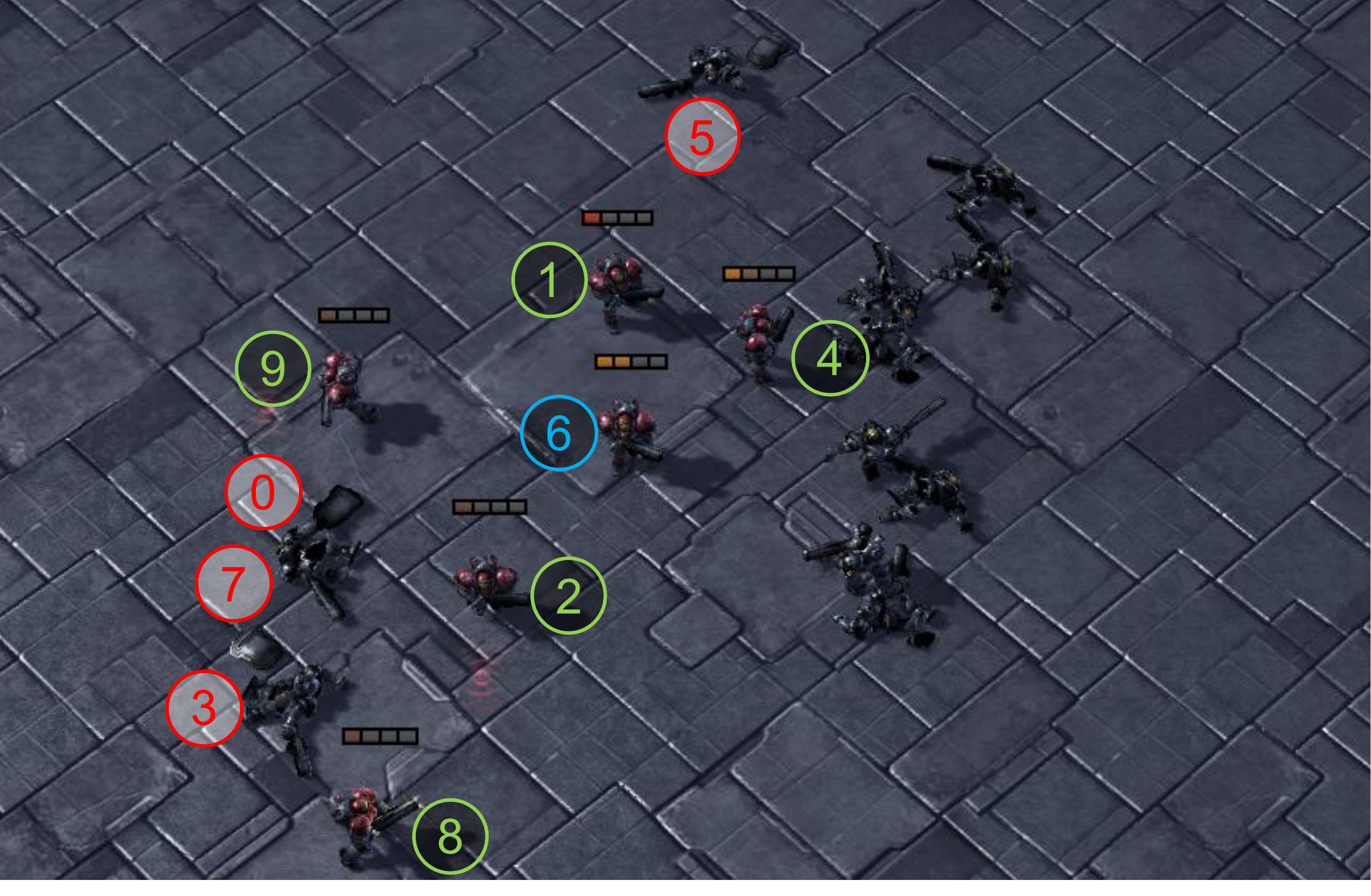}\hfill
\includegraphics[height=0.35\linewidth]{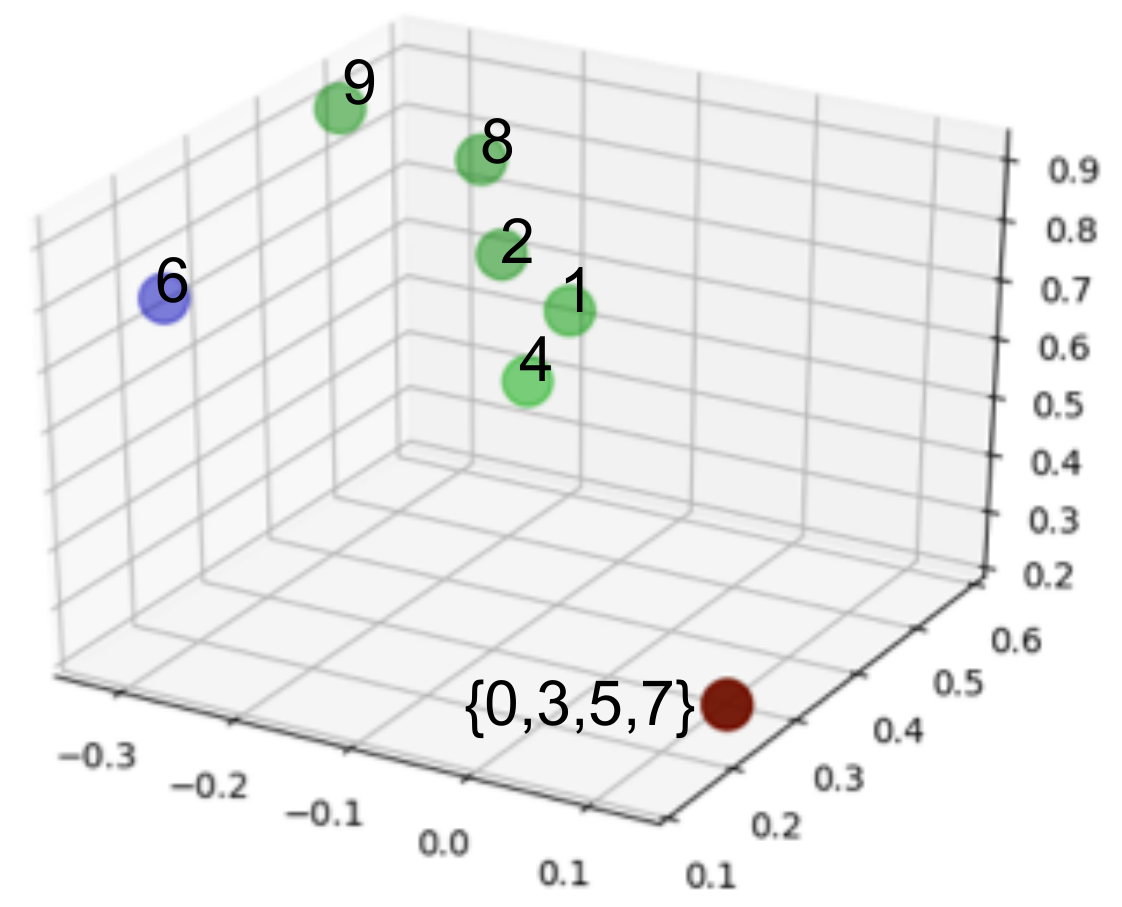}\hfill
\caption{Visualization of our learned role representations at a timestep. The blue agent has the maximum health, while the red ones are dead. The corresponding policy is that $\mathtt{agent\ 6}$ moves towards enemies to take on more firepower, so that more seriously injured agents are protected. Roles can change adaptively and will aggregate according to responsibilities that are compatible with individual characteristics, such as location, agent type, health, etc.}\label{fig:teaser}
\end{figure}

In order to achieve scalability, these deep MARL methods adopt a simple mechanism that all agents share and learn a decentralized value or policy network. However, such simple sharing is often not effective for many complex multi-agent tasks. For example, in Adam Smith's Pin Factory, workers must complete up to eighteen different tasks to create one pin~\cite{smith1937wealth}. In this case, it is a heavy burden for a single shared policy to represent and learn all required skills. On the other hand, it is also unnecessary for each agent to use a distinct policy network, which leads to high learning complexity because some agents often perform similar sub-tasks from time to time. The question is how we can give full play to agents' specialization and dynamic sharing for improving learning efficiency.

A natural concept that comes to mind is the \emph{role}. A role is a comprehensive pattern of behavior, often specialized in some tasks. Agents with similar roles will show similar behaviors, and thus can share their experiences to improve performance. The role theory has been widely studied in economics, sociology, and organization theory. Researchers have also introduced the concept of role into MAS~\cite{becht1999rope, stone1999task, depke2001roles, ferber2003agents, odell2004metamodel, bonjean2014adelfe, Lhaksmana2018role}. In these role-based frameworks, the complexity of agent design is reduced via task decomposition by defining roles associated with responsibilities made up of a set of sub-tasks, so that the policy search space is effectively decomposed~\cite{zhu2008role}. However, these works exploit prior domain knowledge to decompose tasks and predefine the responsibilities of each role, which prevents role-based MAS from being dynamic and adaptive to uncertain environments.

To leverage the benefits of both role-based and learning methods, in this paper, we propose a role-oriented multi-agent reinforcement learning framework (\name). This framework implicitly introduces the role concept into MARL, which serves as an intermediary to enable agents with similar responsibilities to share their learning. We achieve this by ensuring that agents with similar roles have both similar policies and responsibilities. To establish the connection between roles and decentralized policies, \name~conditions agents' policies on individual roles, which are stochastic latent variables determined by agents' local observations. To associate roles with responsibilities, we introduce two regularizers to enable roles to be identifiable by behaviors and specialized in certain sub-tasks. We show how well-formed role representations can be learned via optimizing tractable variational estimations of the proposed regularizers. In this way, our method synergizes role-based and learning methods while avoiding their individual shortcomings -- we provide a flexible and general-purpose mechanism that promotes the \emph{emergence} and \emph{specialization} of roles, which in turn provides an adaptive learning sharing mechanism for efficient multi-agent policy learning.

We test our method on StarCraft II\footnote{StarCraft II are trademarks of Blizzard Entertainment\textsuperscript{TM}.} micromanagement environments~\cite{vinyals2017starcraft, samvelyan2019starcraft}. Results show that our method significantly pushes forward the state of the art of MARL algorithms, by virtue of the adaptive policy sharing among agents with similar roles. Visualization of the role representations in both homogeneous and heterogeneous agent teams demonstrates that the learned roles can adapt automatically in dynamic environments, and that agents with similar responsibilities have similar roles. In addition, the emergence and evolution process of roles is shown, highlighting the connection between role-driven sub-task specialization and improvement of team efficiency in our framework. These results provide a new perspective in understanding and promoting the emergence of cooperation among agents.

%% file: 2-Preliminaries.tex
\section{Background}
In our work, we consider a fully cooperative multi-agent task that can be modelled by a Dec-POMDP~\cite{oliehoek2016concise} $G\shorte\langle I, S, A, P, R, \Omega, O, n, \gamma\rangle$, where $A$ is the finite action set, $I$ is the finite set of $n$ agents, $\gamma\in[0, 1)$ is the discount factor, and $s\in S$ is the true state of the environment. We consider partially observable settings and agent $i$ only has access to an observation $o_i\in \Omega$ drawn according to the observation function $O(s, i)$. Each agent has a history $\tau_i\in \Tau\equiv(\Omega\times A)^*$. At each timestep, each agent $i$ selects an action $a_i\in A$, forming a joint action $\va$ $\in A^n$, leading to next state $s'$ according to the transition function $P(s'|s, \va)$ and a shared reward $r=R(s,\va)$ for each agent. The joint policy $\bm{\pi}$ induces a joint action-value function: $Q_{tot}^{\bm{\pi}}(s$,$ \va)$=$\mathbb{E}_{s_{0:\infty},\va_{0:\infty}}[\sum_{t=0}^\infty \gamma^{t}r_t|$ $s_0\shorte s,\va_0\shorte \va,\bm{\pi}]$. 

To effectively learn policies for agents, the paradigm of centralized training with decentralized execution (CTDE)~\cite{foerster2016learning, foerster2018counterfactual, wang2020influence} has recently attracted attention from deep MARL to deal with non-stationarity while learning decentralized policies. One of the promising ways to exploit the CTDE paradigm is value function decomposition~\cite{sunehag2018value, rashid2018qmix, son2019qtran, wang2020learning}, which learns a decentralized utility function for each agent and uses a mixing network to combine these local utilities into a global action value. To achieve learning scalability, existing CTDE methods typically learn a shared local value or policy network for agents. However, this simple sharing mechanism is often not sufficient for learning complex tasks, where diverse responsibilities or skills are required to achieve goals. In this paper, we develop a novel role-based MARL framework to address this challenge. This framework achieves efficient shared learning while allowing agents to learn sufficiently diverse skills. 


%% file: 3-Methods.tex
\section{Method}

\begin{figure}
    \centering
    \includegraphics[width=\linewidth]{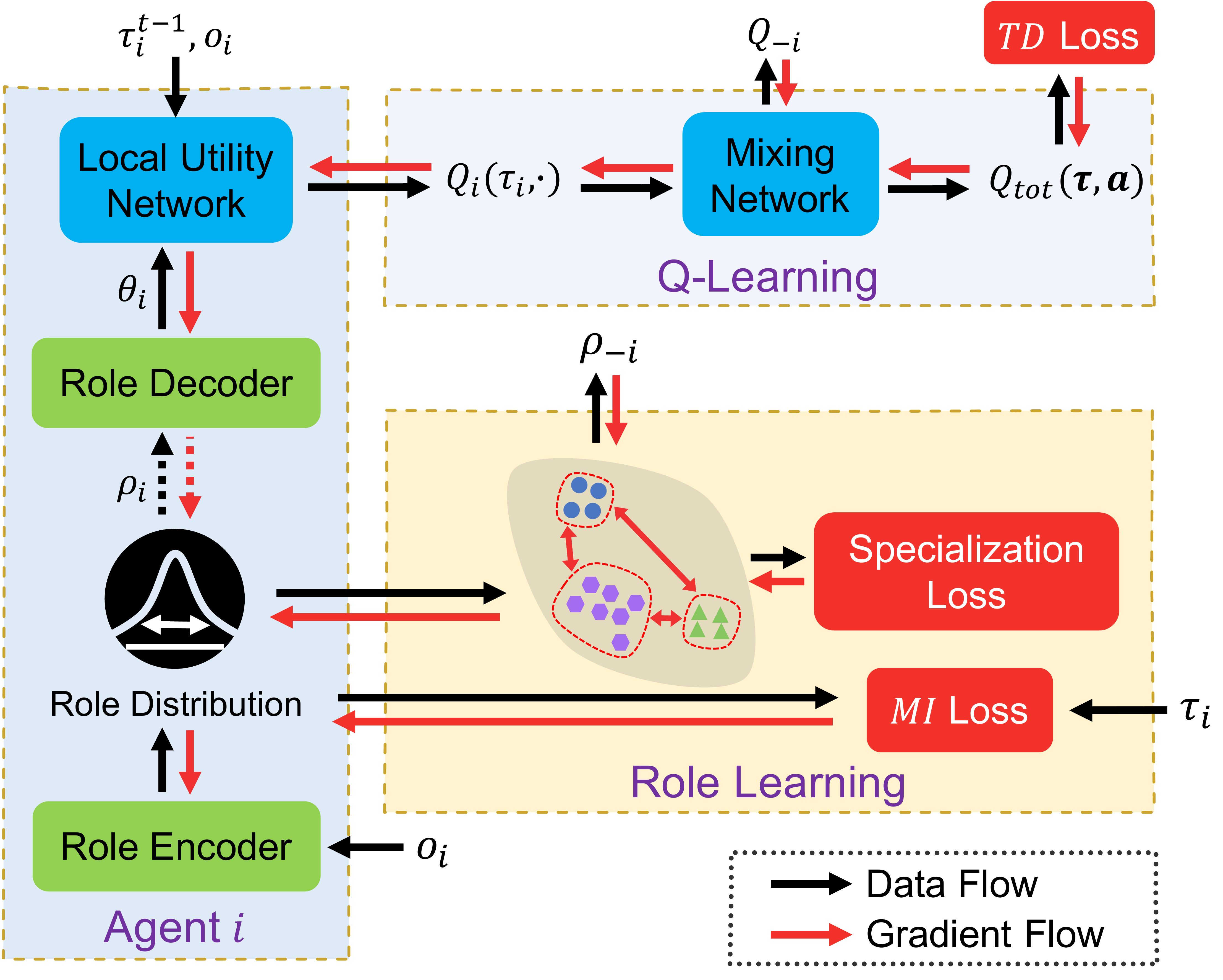}
    \caption{Schematics of our approach. The role encoder generates a role embedding distribution, from which a role is sampled and serves as the input to the role decoder. The role decoder generates the parameters of the local utility network. Local utilities are fed into a mixing network to get an estimation of the global action value. We propose two learning objectives to learn specialized and identifiable roles. The framework can be trained in an end-to-end manner.}
    \label{fig:framework}
\end{figure}

In this section, we will present a novel role-oriented MARL framework (\name) that introduces the role concept into MARL and enables adaptive shared learning among agents. \name~adopts the CTDE paradigm. As shown in Fig.~\ref{fig:framework}, it learns local Q-value functions for agents, which are fed into a mixing network to compute a global TD loss for centralized training. During the execution, the mixing network will be removed, and each agent will act based on its local policy derived from its value function. Agents' value functions or policies are dependent on their roles, each of which is responsible for performing similar automatically identified sub-tasks. To enable efficient and effective shared learning among agents with similar behaviors, \name~will automatically learn roles that are: 


i) \textbf{Dynamic}: An agent's role can automatically adapt to the dynamics of the environment;

ii) \textbf{Identifiable}: The role of an agent contains enough information about its behaviors; 

iii) \textbf{Specialized}: Agents with similar roles are expected to specialize in similar sub-tasks.

Formally, each agent $i$ has a local utility function (or an individual policy), whose parameters $\theta_i$ are conditioned on its role $\rho_i$. To learn roles with desired properties, we encode roles in a stochastic embedding space, and the role of agent $i$, $\rho_i$, is drawn from a multivariate Gaussian distribution $\mathcal{N}(\bm{\mu}_{\rho_i}, \bm{\sigma}_{\rho_i})$. To enable the \textbf{dynamic} property, \name~conditions an agent's role on its local observations, and uses a trainable neural network $f$ to learn the parameters of the Gaussian distribution of the role: 
\begin{equation}
\begin{aligned}
(\bm{\mu}_{\rho_i}, \bm{\sigma}_{\rho_i}) &= f(o_i; \theta_\rho), \\
\rho_i &\sim \mathcal{N}(\bm{\mu}_{\rho_i}, \bm{\sigma}_{\rho_i}),
\end{aligned}
\end{equation}
where $\theta_\rho$ are parameters of $f$. The sampled role $\rho_i$ is then fed into a hyper-network $g(\rho_i; \theta_h)$ parameterized by $\theta_h$ to generate the parameters for the individual policy, $\theta_i$. We call $f$ the \emph{role encoder} and $g$ the \emph{role decoder}. In the next two sub-sections, we will describe two regularizers for learning identifiable and specialized roles. 

\subsection{Identifiable Roles}\label{sec:indentifiable_roles}
Introducing latent role embedding and conditioning individual policies on this embedding does not automatically generate roles with desired properties. Intuitively, conditioning roles on local observations enables roles to be responsive to the changes in the environment. This design enables \name~to be adaptive to dynamic environments but may cause roles to change quickly, making learning unstable. For addressing this problem, we expect roles to be temporally stable. To this end, we propose to learn roles that are identifiable by agents' long term behaviors, which can be achieved by maximizing $I(\tau_i; \rho_i | o_i)$, the conditional mutual information between the individual trajectory and the role given the current observation.

However, estimating and maximizing mutual information is often intractable. Drawing inspiration from the literature of variational inference~\cite{wainwright2008graphical, alemi2017deep}, we introduce a variational posterior estimator to derive a tractable lower bound for the mutual information objective (the proof is deferred to Appendix A.1):
\begin{equation}
   I(\rho^t_i; \tau^{t-1}_i | o^t_i) \ge  \mathbb{E}_{\rho^t_i, \tau^{t-1}_i, o^t_i}\left[\log\frac{q_\xi(\rho^t_i | \tau^{t-1}_i, o^t_i)}{p(\rho^t_i | o^t_i)}\right],
   \label{equ:mi}
\end{equation}
where $\tau^{t-1}_i=(o_i^0, a_i^0, \cdots, o_i^{t-1}, a_i^{t-1})$, $q_\xi$ is the variational estimator parameterised with $\xi$. For $q_\xi$, we use a GRU~\cite{cho2014learning} to encode an agent's history of observations and actions, and call it the \emph{trajectory encoder}. The lower bound in Eq.~\ref{equ:mi} can be further rewritten as a loss function to be minimized:
\begin{equation}
    \mathcal{L}_I(\theta_\rho, \xi) = \mathbb{E}_{(\tau^{t\shortn 1}_i, o^t_i)\sim D}\left[\KL[p(\rho^{t}_i | o^t_i)\| q_\xi(\rho^{t}_i | \tau^{t\shortn 1}_i, o^t_i)]\right],
\end{equation}
where $\mathcal{D}$ is a replay buffer, and $\KL[\cdot\|\cdot]$ is the KL divergence operator. The detailed derivation can be found in Appendix A.1.

\begin{figure*}
\centering
\includegraphics[height=0.17\linewidth]{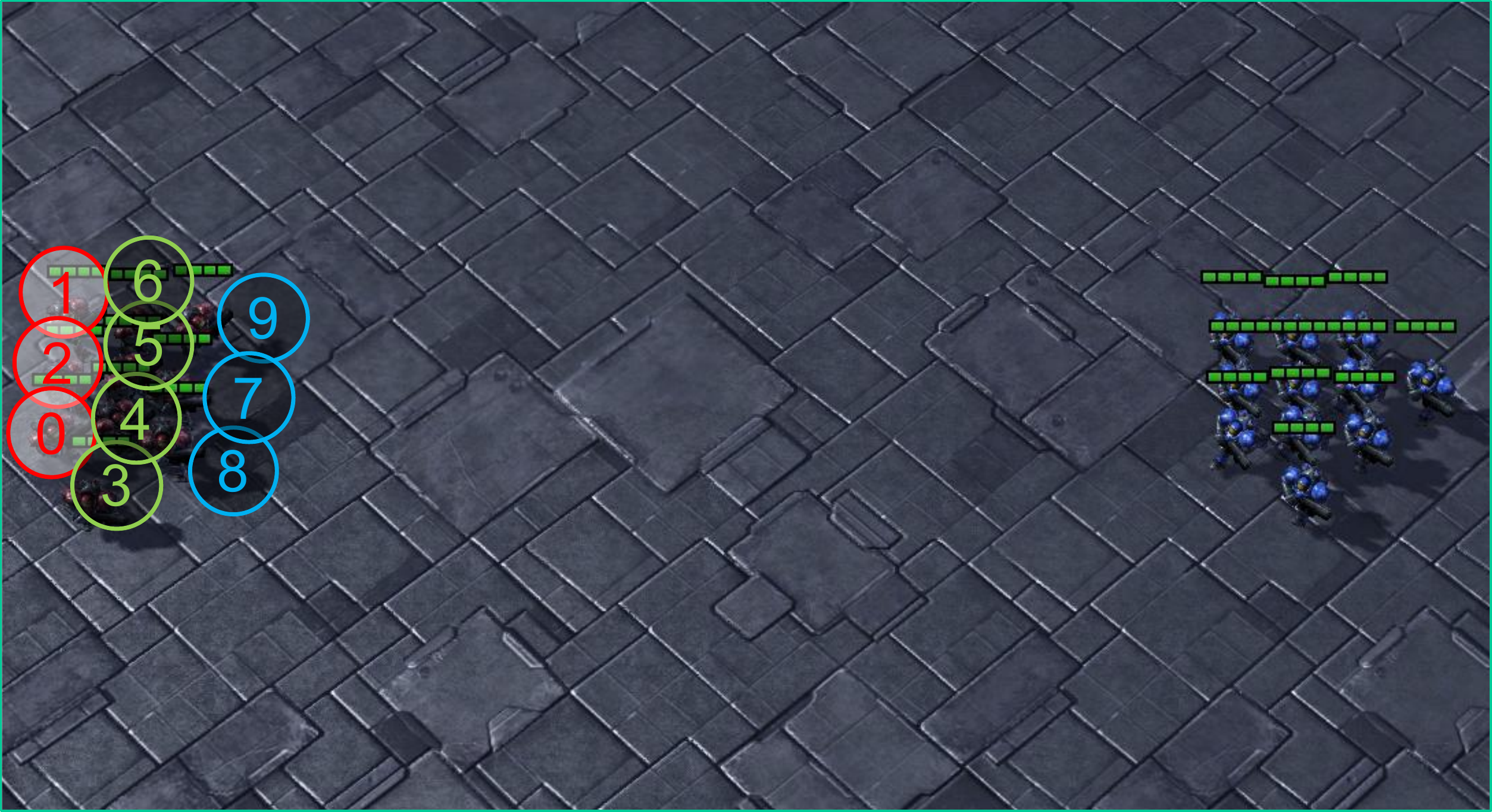}\hfill
\includegraphics[height=0.17\linewidth]{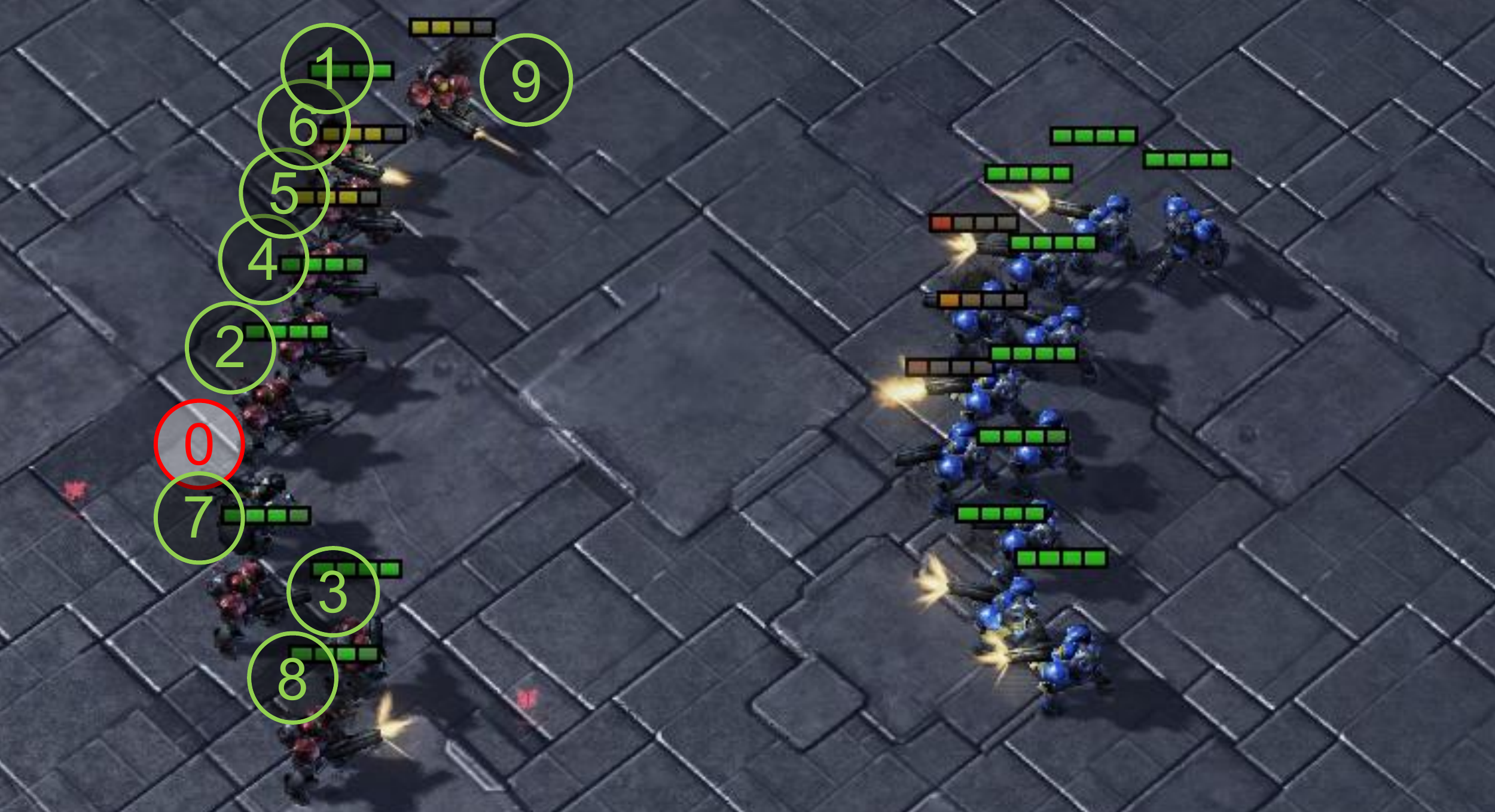}\hfill
\includegraphics[height=0.17\linewidth]{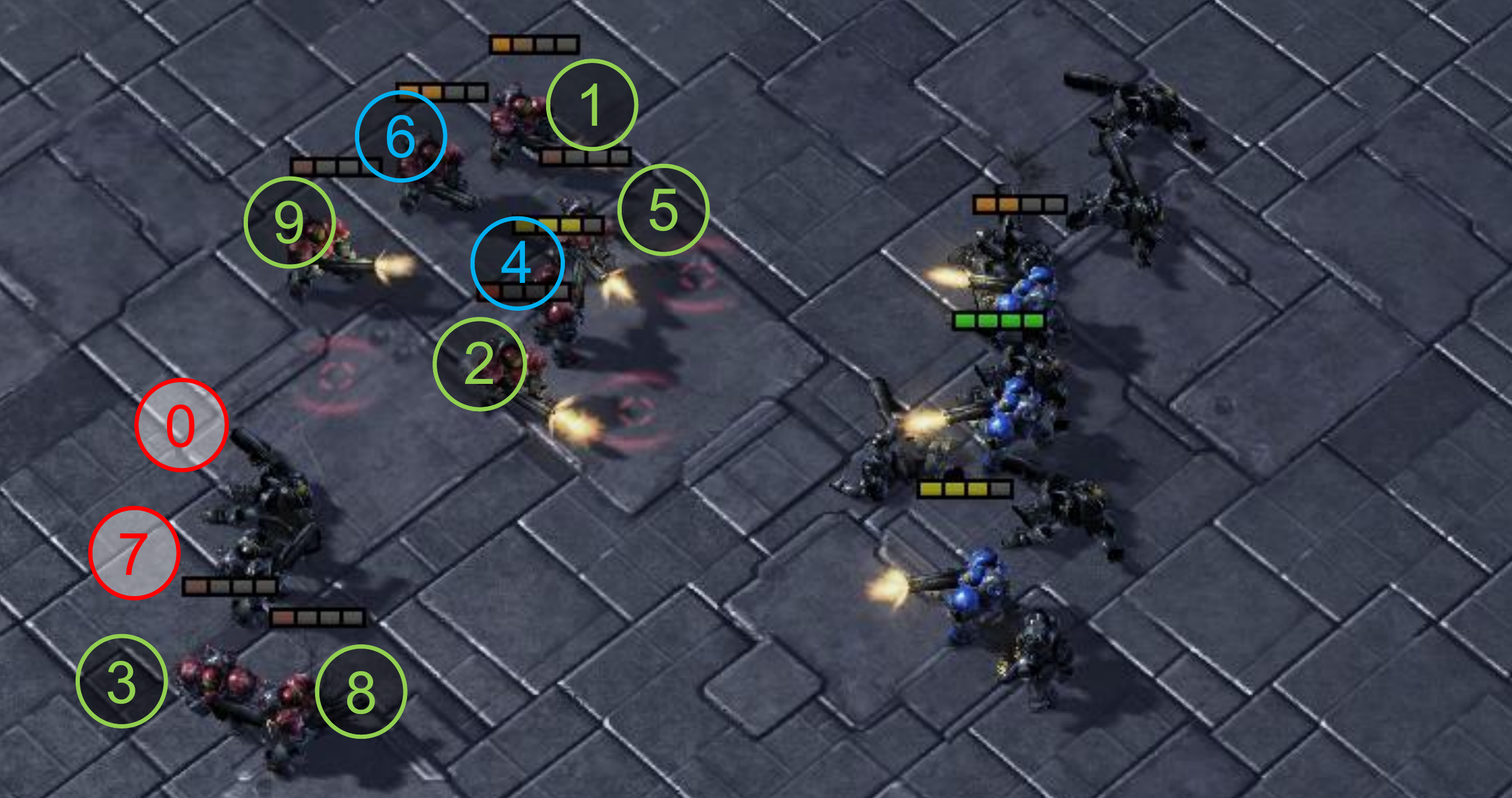}\\ 
\subfigure[$t$=$1$, roles are characterized by agents' positions.]{\includegraphics[width=0.32\linewidth]{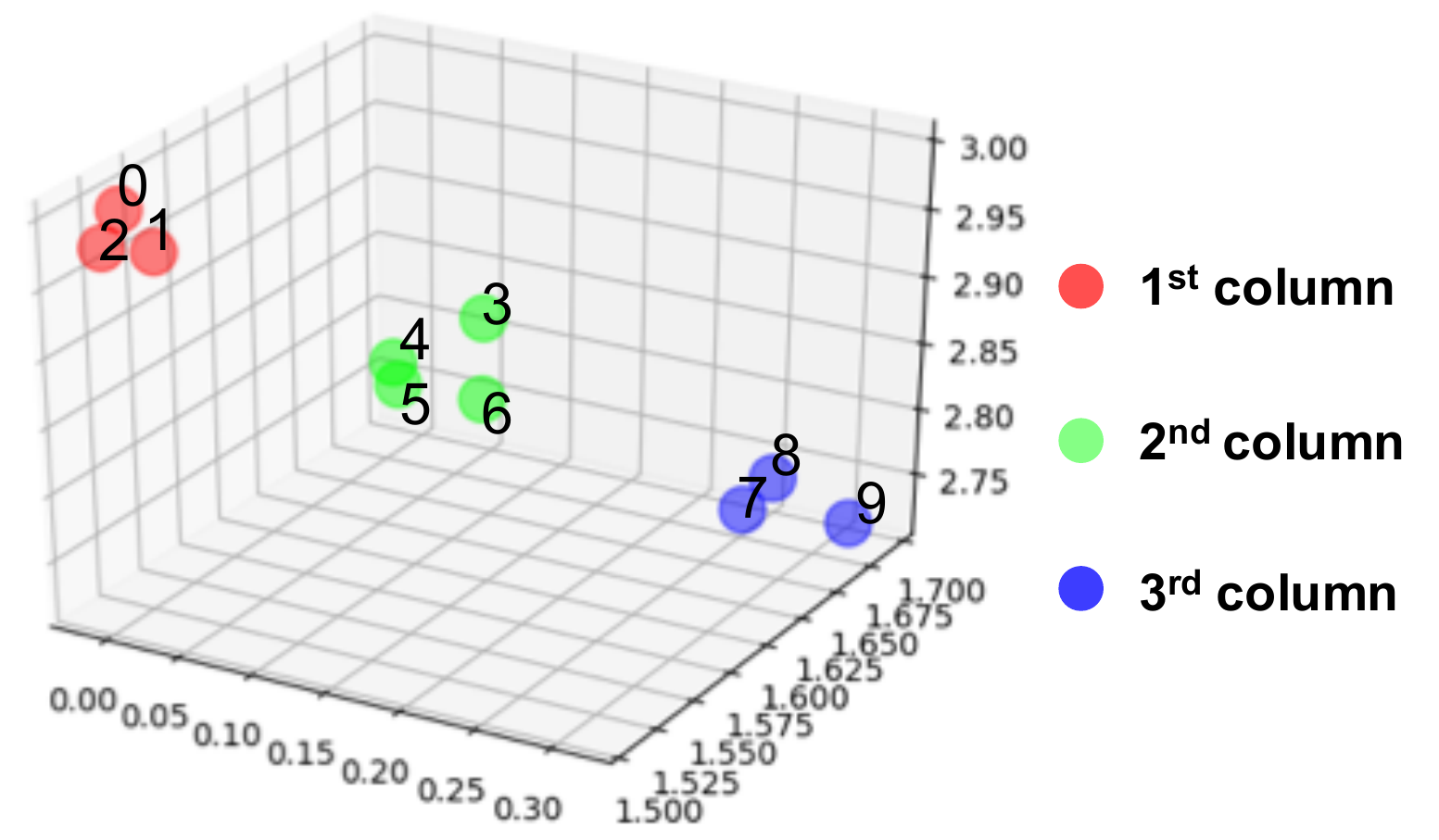}}\hfill
\subfigure[$t$=$8$, roles are characterized by agents' remaining hit points.]{\includegraphics[width=0.32\linewidth]{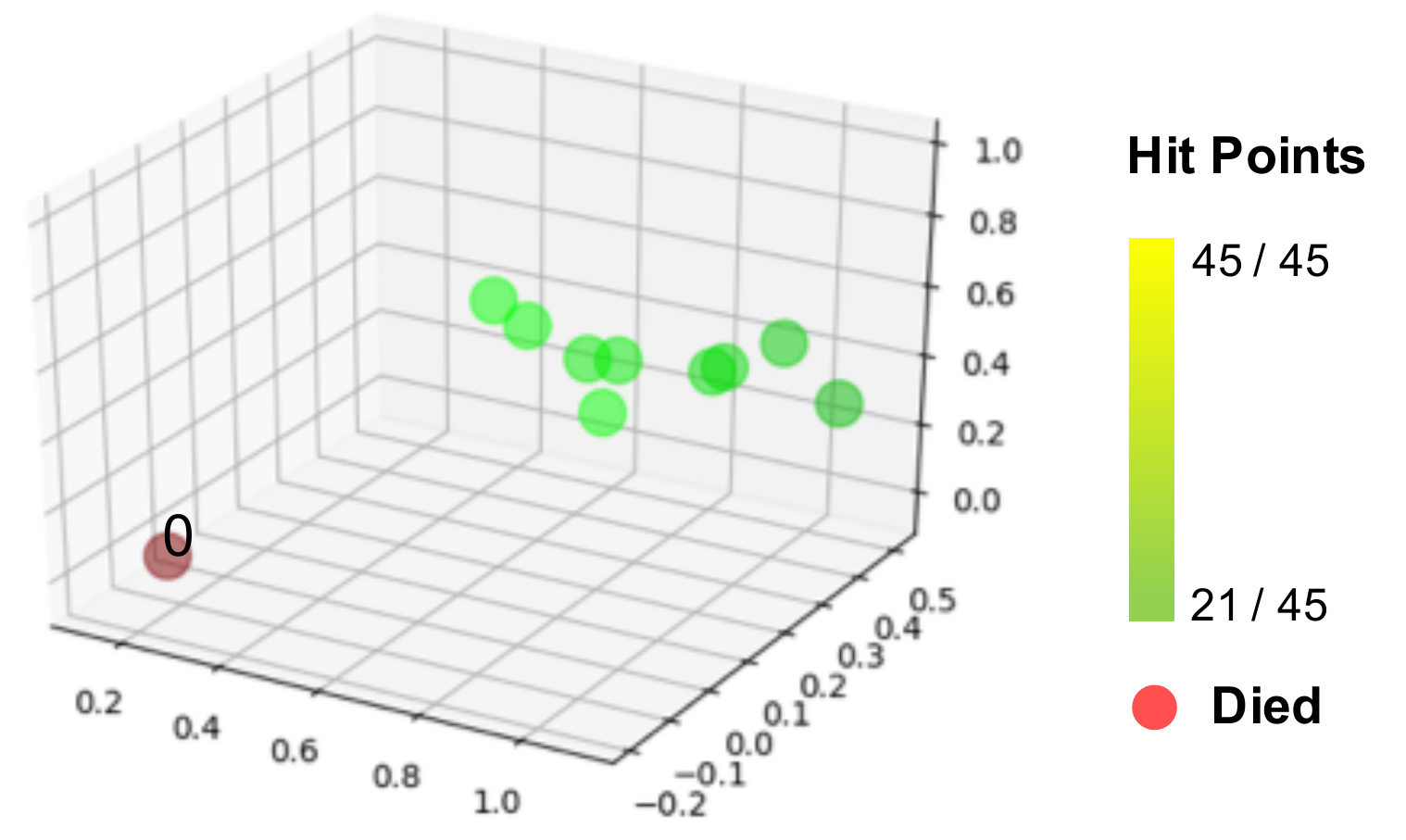}}\hfill
\subfigure[$t$=$19$, roles are characterized by agents' remaining hit points and positions.]{\includegraphics[width=0.32\linewidth]{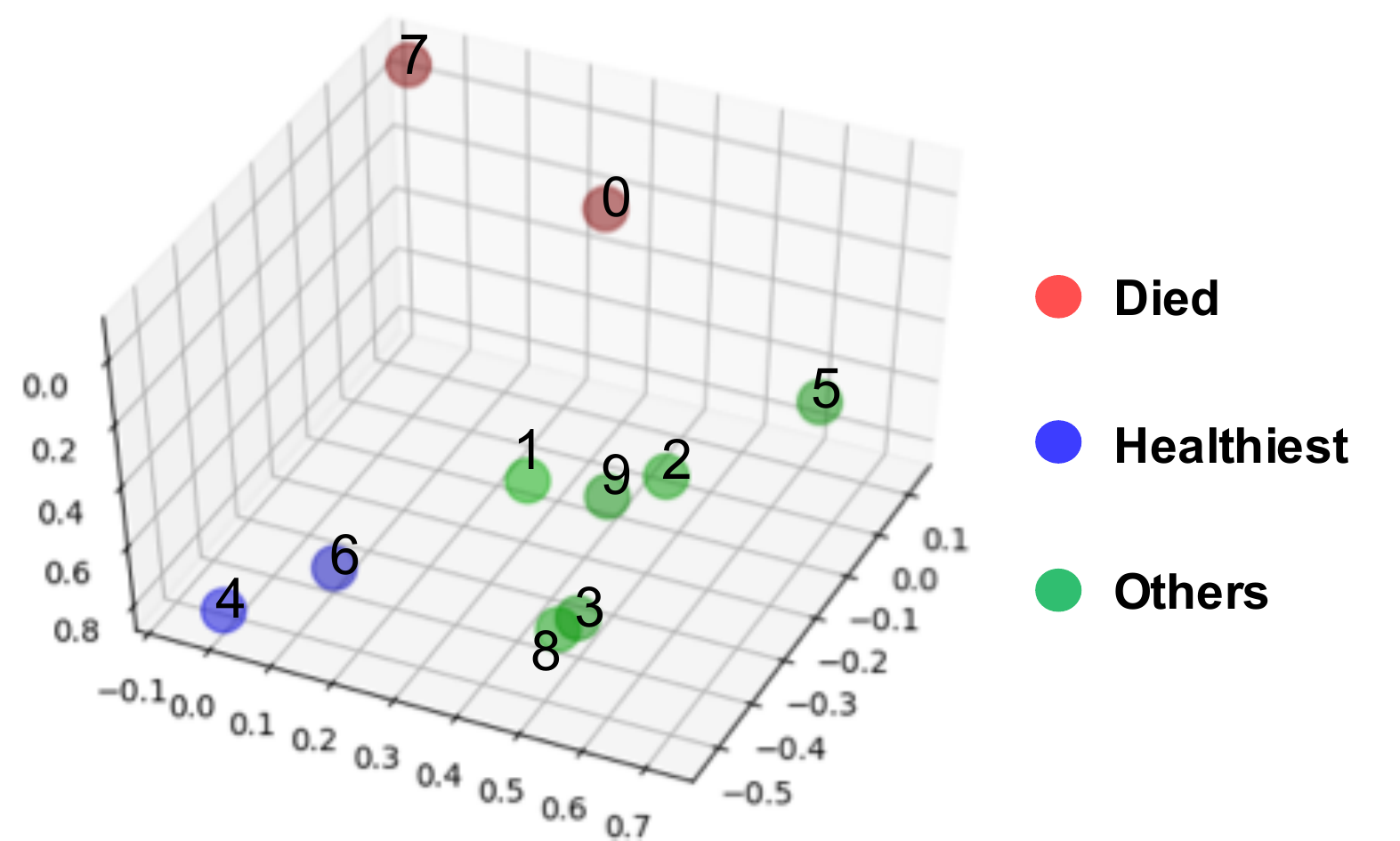}}
\caption{Dynamic role adaptation during an episode (means of the role distributions, $\bm{\mu}_{\rho_i}$, are shown, without using any dimensionality reduction techniques). The \emph{role encoder} learns to focus on different parts of observations according to the automatically discovered demands of the task. The role-induced strategy helps (a) quickly form the offensive arc when $t$=$1$; (b) protect injured agents when $t$=$8$; (c) protect dying agents and alternate fire when $t=19$.}\label{fig:dynamic_role-10m_vs_11m}
\end{figure*}

\subsection{Specialized Roles}\label{sec:specialized_roles}
The formulation so far does not promote sub-task specialization, which is the critical component to share learning and improve efficiency in multi-agent systems. Minimizing $\mathcal{L}_I$ enables roles to contain enough information about long-term behaviors but does not explicitly ensure agents with similar behaviors to have similar role embeddings.

For learning specialized roles, we define another role-learning regularizer. Intuitively, to encourage sub-task specialization, for any two agents, we expect that either they have similar roles or they have quite different behaviors. However, it is usually unclear which agents will have similar roles during the process of role emergence, and the similarity between behaviors is not straightforward to measure. 

Since roles have enough information about the behaviors (achieved by minimizing $\mathcal{L}_I$), to encourage two agents $i$ and $j$ to have similar roles, we can maximize $I(\rho_i; \tau_j)$, the mutual information between the role of agent $i$ and the trajectory of agent $j$. However, we do not know which agents will have similar roles, and directly optimizing this objective for all pairs of agents will result in all agents having the same role, and, correspondingly, the same policy, which will limit system performance. To settle this issue, we introduce a dissimilarity model $d_{\phi}: \Tau\times\Tau\rightarrow\mathbb{R}$, a trainable neural network taking two trajectories as input, and seek to maximize $I(\rho_i; \tau_j) + d_\phi(\tau_i, \tau_j)$ while minimizing the number of non-zero elements in the matrix $D_\phi=(d_{ij})$. Here, $d_{ij}=d_\phi(\tau_i, \tau_j)$ is the estimated dissimilarity between trajectories of agent $i$ and $j$. Such formulation makes sure that dissimilarity $d$ is high only when mutual information $I$ is low, so that the set of learned roles is compact but diverse, which help solve the given task efficiently. Formally, the following learning objective encourages sub-task specialization:
\begin{align*}
    &\underset{{\theta_\rho, \xi, \phi}}{\text{minimize}} \ \ \ \|D_\phi^t\|_{2,0} \stepcounter{equation}\tag{\theequation} \\
    &\text{subject to} \ \ \ I(\rho^{t}_i; \tau^{t-1}_j|o^t_j) + d_\phi(\tau^{t-1}_i, \tau^{t-1}_j) > U, \forall i\ne j,
\end{align*}
where $U$ controls the compactness of the role representation. In practice, we separately carry out min-max normalization on $I$ and $d$ to scale their values to $[0,1]$ and set $U$ to 1. Relaxing the matrix norm $\|\cdot\|_{2,0}$ with the Frobenius norm, we can get the optimization objective for minimizing:
\begin{equation}
\begin{aligned}
\|D_\phi^t\|_{F} - \sum_{i\ne j}\min\{I(\rho^{t}_i; \tau^{t-1}_j|o^t_j) + d_\phi(\tau^{t-1}_i, \tau^{t-1}_j), U\},
\end{aligned}
\end{equation}
However, as estimating and optimizing the mutual information term are intractable, we use the variational posterior estimator introduced in Sec.~\ref{sec:indentifiable_roles} to construct an upper bound, serving as the second regularizer of \name:
\begin{align*}
    \mathcal{L}_D&(\theta_\rho, \phi, \xi) = \mathbb{E}_{(\bm{\tau}^{t-1}, \bm{o}^t)\sim\mathcal{D}, \bm{\rho}^t\sim p(\bm{\rho}^t | \bm{o}^t)}\big[\|D_\phi^t\|_{F} \stepcounter{equation}\tag{\theequation} \\ 
        & - \sum_{i\ne j}\min\{q_\xi(\rho^{t}_i| \tau^{t-1}_j, o^t_j) + d_\phi(\tau^{t-1}_i, \tau^{t-1}_j), U\}\big]
\end{align*}
where $\mathcal{D}$ is the replay buffer, $\bm{\tau}^{t-1}$ is the joint trajectory, $\bm{o}^{t}$ is the joint observation, and $\bm{\rho}^{t}=\langle \rho_1^t, \rho_2^t, \cdots, \rho_n^t\rangle$. A detailed derivation can be found in Appendix A.2.
\begin{figure*}
\includegraphics[width=\linewidth]{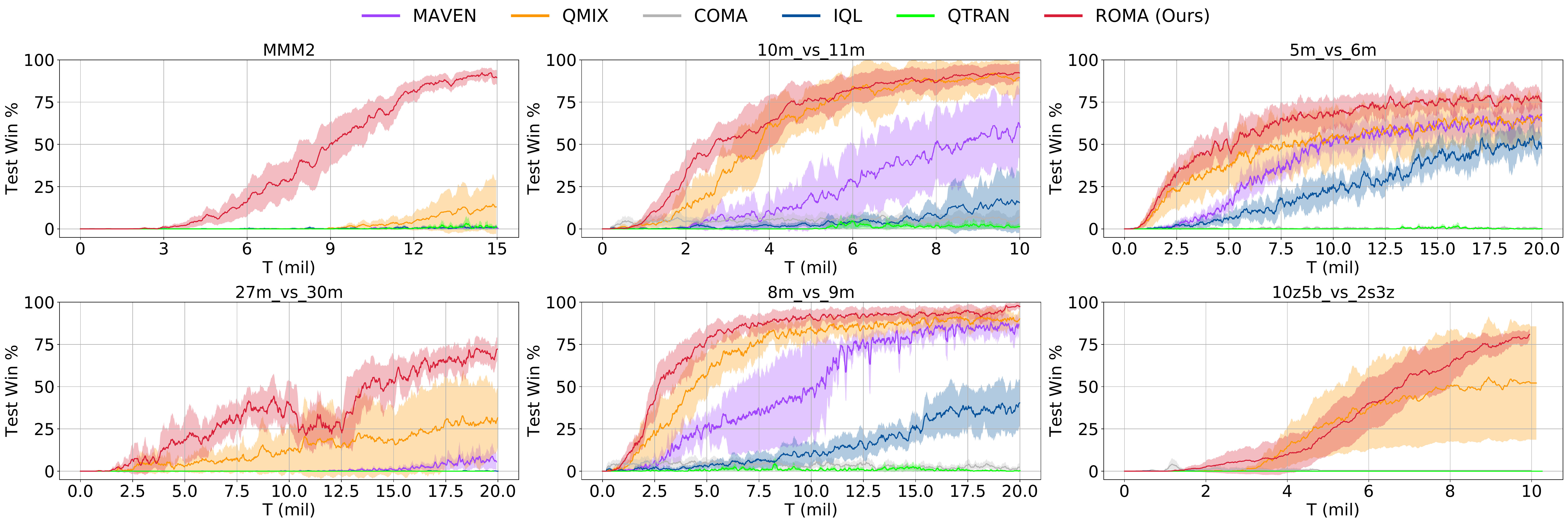}
\caption{Comparison of our method against baseline algorithms. Results for more maps can be found in Appendix C.1.}\label{fig:performance-baselines}
\end{figure*}

\subsection{Overall Optimization Objective}
We have introduced optimization objectives for learning roles to be identifiable and and specialized. Apart from these regularizers, all the parameters in the framework are updated by gradients induced by the standard TD loss of reinforcement learning. As shown in Fig.~\ref{fig:framework}, to compute the global TD loss, individual utilities are fed into a mixing network whose output is the estimation of global action-value $Q_{tot}$. In this paper, our \name~implementation uses the mixing network introduced by QMIX~\cite{rashid2018qmix} (see Appendix D) for its monotonic approximation, but it can be easily replaced by other mixing methods. The parameters of the mixing network are conditioned on the global state $s$ and are generated by a hyper-net parameterized by $\theta_m$. Therefore, the final learning objective of \name~is:
\begin{equation}
    \mathcal{L}(\theta) = \mathcal{L}_{TD}(\theta) +\lambda_I\mathcal{L}_I(\theta_\rho, \xi) + \lambda_D\mathcal{L}_D(\theta_\rho, \xi, \phi),
\end{equation}
where $\theta=(\theta_\rho, \xi, \phi, \theta_h, \theta_m)$, $\lambda_I$ and $\lambda_D$ are scaling factors, and $\mathcal{L}_{TD}(\theta)$ = $[r + \gamma \max_{\va'} Q_{tot}(s', \bm{a'}; \theta^-)$-$Q_{tot}$($s, \bm{a}; \theta$)$]^2$ ($\theta^-$ are the parameters of a periodically updated target network). In our centralized training with decentralized execution framework, only the role encoder, the role decoder, and the individual utility networks are used when execution.


%% file: 4-RelatedWorks.tex
\section{Related Works}

The emergence of role has been documented in many natural systems, such as bees~\cite{jeanson2005emergence}, ants~\cite{gordon1996organization}, and humans~\cite{butler2012condensed}. In these systems, the role is closely related to the division of labor and is crucial to the improvement of labor efficiency. Many multi-agent systems are inspired by these natural systems. They decompose the task, make agents with the same role specialize in certain sub-tasks, and thus reduce the design complexity~\cite{wooldridge2000gaia, omicini2000soda, padgham2002prometheus, pavon2003agent, cossentino2005passi, zhu2008role, spanoudakis2010using, deloach2010mase, bonjean2014adelfe}. These methodologies are designed for tasks with a clear structure, such as software engineering~\cite{bresciani2004tropos}. Therefore, they tend to use predefined roles and associated responsibilities~\cite{ Lhaksmana2018role}. In contrast, we focus on how to implicitly introduce the concept of roles into general multi-agent sequential decision making under dynamic and uncertain environments.

Deep multi-agent reinforcement learning has witnessed vigorous progress in recent years. COMA~\citep{foerster2018counterfactual}, MADDPG~\citep{lowe2017multi}, PR2~\citep{wen2019probabilistic}, and MAAC~\cite{iqbal2019actor} explore multi-agent policy gradients. Another line of research focuses on value-based multi-agent RL, and value-function factorization is the most popular method. VDN~\citep{sunehag2018value}, QMIX~\citep{rashid2018qmix}, and QTRAN~\citep{son2019qtran} have progressively enlarged the family of functions that can be represented by the mixing network. NDQ~\cite{wang2020learning} proposes nearly decomposable value functions to address the miscoordination problem in learning fully decentralized value functions. Emergence is a topic with increasing interest in deep MARL. Works on the emergence of communication~\cite{foerster2016learning, lazaridou2017multi, das2017learning, mordatch2018emergence, wang2020learning, kang2020incorporating}, the emergence of fairness~\cite{jiang2019learning}, and the emergence of tool usage~\cite{baker2020emergent} provide a deep learning perspective in understanding both natural and artificial multi-agent systems.

To learn diverse and identifiable roles, we propose to optimize the mutual information between individual roles and trajectories. A recent work studying multi-agent exploration, MAVEN~\cite{mahajan2019maven}, uses a similar objective. Different from \name, MAVEN aims at committed exploration. This difference in high-level purpose leads to many technical distinctions. First, MAVEN optimizes the mutual information between the joint trajectory and a latent variable conditioned on a Gaussian or uniform random variable to encourage diverse joint trajectory. Second, apart from the mutual information objective, we propose a novel regularizer to learn specialized roles, while MAVEN adopts a hierarchical structure and encourages the latent variable to help get more environmental rewards. We empirically compare \name~with MAVEN in Sec.~\ref{sec:exp}. More related works will be discussed in Appendix D.

%% file: 5-Experiments.tex
\begin{figure*}
\includegraphics[width=\linewidth]{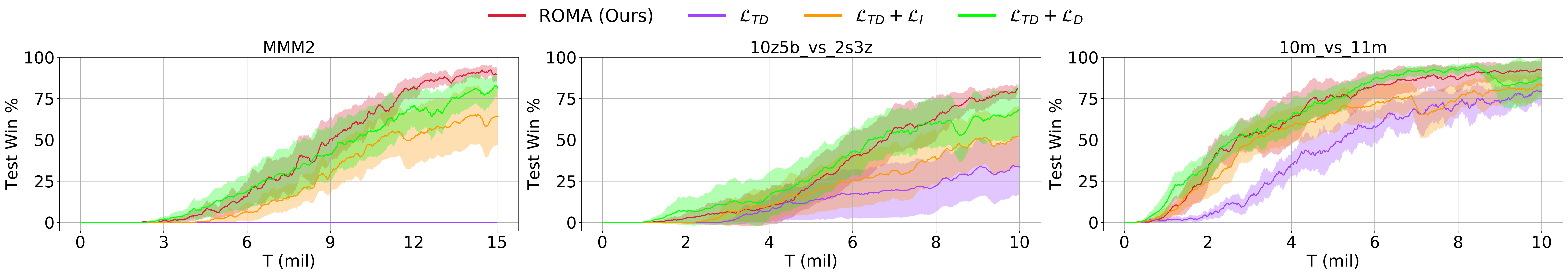}
\caption{Ablation studies regarding the two role-learning losses.}\label{fig:performance-ablations-losses}
\end{figure*}
\begin{figure*}
\includegraphics[width=\linewidth]{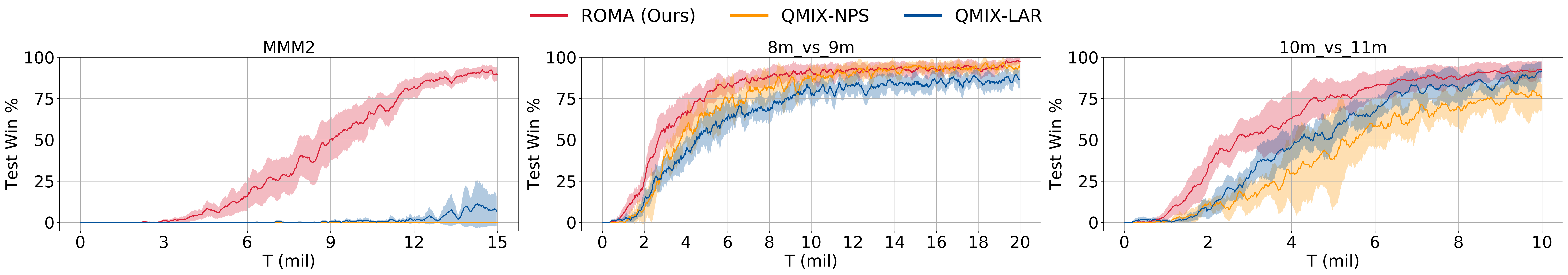}
\caption{Comparison of our method against ablations.}\label{fig:performance-ablations}
\end{figure*}

\section{Experiments}\label{sec:exp}

Our experiments aim to answer the following questions: (1) Whether the learned roles can automatically adapt in dynamic environments? (Sec.~\ref{sec:exp-dynamic_role}.) (2) Can our method promote sub-task specialization? That is, agents with similar responsibilities have similar role embedding representations, while agents with different responsibilities have role embedding representations far from each other. (Sec.~\ref{sec:exp-dynamic_role},~\ref{sec:exp-various_roles}.) (3) Can such sub-task specialization improve the performance of multi-agent reinforcement learning algorithms? (Sec.~\ref{sec:exp-performance}.) (4) How do roles evolve during training, and how do they influence team performance? (Sec.~\ref{sec:exp-role_evolution}.) (5) Can the dissimilarity model $d_{\phi}$ learn to measure the dissimilarity between agents' trajectories? (Sec.~\ref{sec:exp-role_evolution}.) Videos\footnote{\url{https://sites.google.com/view/romarl/}} of our experiments and the code\footnote{\url{https://github.com/TonghanWang/ROMA}} are available online.

\textbf{Baselines}\ \ We compare our methods with various baselines shown in Table~\ref{tab:baselines}. In particular, we carry out the following ablation studies: (i) We separately omit each (or both) of the two role-learning objectives ($\mathcal{L}_I$ and $\mathcal{L}_D$) while leaving the other parts of ROMA unchanged. These three ablations are designed to highlight the contribution of each of the proposed regularizers. (ii) QMIX-NPS. The same as QMIX~\cite{rashid2018qmix}, but agents do not share parameters. Our method achieves adaptive learning sharing, and comparison against QMIX (parameters are shared among agents) and QMIX-NPS tests whether this flexibility can improve learning efficiency. (iii) QMIX-LAR, QMIX with a similar number of parameters with our framework, which can test whether the superiority of our method comes from the increase in the number of parameters.

We carry out a grid search over the loss coefficients $\lambda_I$ and $\lambda_D$, and fix them at $10^{-4}$ and $10^{-2}$, respectively, across all the experiments. The dimensionality of latent role space is set to 3, so we did not use any dimensionality reduction techniques when visualizing the role embedding representations. Other hyperparameters are also fixed in our experiments, which are listed in Appendix B.1. For \name, We use elementary network structures (fully-connected networks or GRU) for the role encoder, role decoder, and trajectory encoder. The details of the architecture of our method and baselines can be found in Appendix B.
\begin{table} [t]
    \caption{Baseline algorithms.}
    \label{tab:baselines}
    \centering
    \begin{tabular}{crcrcr}
        \toprule
        \multicolumn{2}{c}{} &
        \multicolumn{2}{l}{Alg.} &
        \multicolumn{2}{l}{Description} \\
        \cmidrule(lr){1-2}
        \cmidrule(lr){3-4}
        \cmidrule(lr){5-6}
        
        \multicolumn{2}{c}{\multirow{5}{*}{\makecell{Related\\ Works}}} &  \multicolumn{2}{l}{IQL} & \multicolumn{2}{l}{Independent Q-learning} \\
        \multicolumn{2}{c}{} & \multicolumn{2}{l}{COMA} & \multicolumn{2}{l}{Foerster et~al.~\yrcite{foerster2018counterfactual}} \\
        \multicolumn{2}{c}{} & \multicolumn{2}{l}{QMIX} & \multicolumn{2}{l}{Rashid et~al.~\yrcite{rashid2018qmix}} \\
        \multicolumn{2}{c}{} & \multicolumn{2}{l}{QTRAN} & \multicolumn{2}{l}{Son et~al.~\yrcite{son2019qtran}} \\
        \multicolumn{2}{c}{} & \multicolumn{2}{l}{MAVEN} & \multicolumn{2}{l}{Mahajan et~al.~\yrcite{mahajan2019maven}} \\
        
        \cmidrule(lr){1-2}
        \cmidrule(lr){3-4}
        \cmidrule(lr){5-6}
        
        \multicolumn{2}{c}{\multirow{7}{*}{\makecell{Abla-\\tions}}} & \multicolumn{2}{l}{$\mathcal{L}_{TD}$} & \multicolumn{2}{l}{ROMA without $\mathcal{L}_I$ and $\mathcal{L}_D$} \\
        \multicolumn{2}{c}{} & \multicolumn{2}{l}{$\mathcal{L}_{TD}+\mathcal{L}_{I}$} & \multicolumn{2}{l}{ROMA without $\mathcal{L}_{D}$} \\
        \multicolumn{2}{c}{} & \multicolumn{2}{l}{$\mathcal{L}_{TD}+\mathcal{L}_{D}$} & \multicolumn{2}{l}{ROMA without $\mathcal{L}_{I}$} \\
        \multicolumn{2}{c}{} & \multicolumn{2}{l}{QMIX-NPS} & \multicolumn{2}{l}{\makecell[l]{QMIX without parameter \\sharing among agents}} \\
        \multicolumn{2}{c}{} & \multicolumn{2}{l}{QMIX-LAR} & \multicolumn{2}{l}{\makecell[l]{QMIX with similar number \\of parameters with ROMA}} \\
        \toprule
    \end{tabular}
\end{table}

\begin{figure*}
\centering
\includegraphics[width=0.32\linewidth]{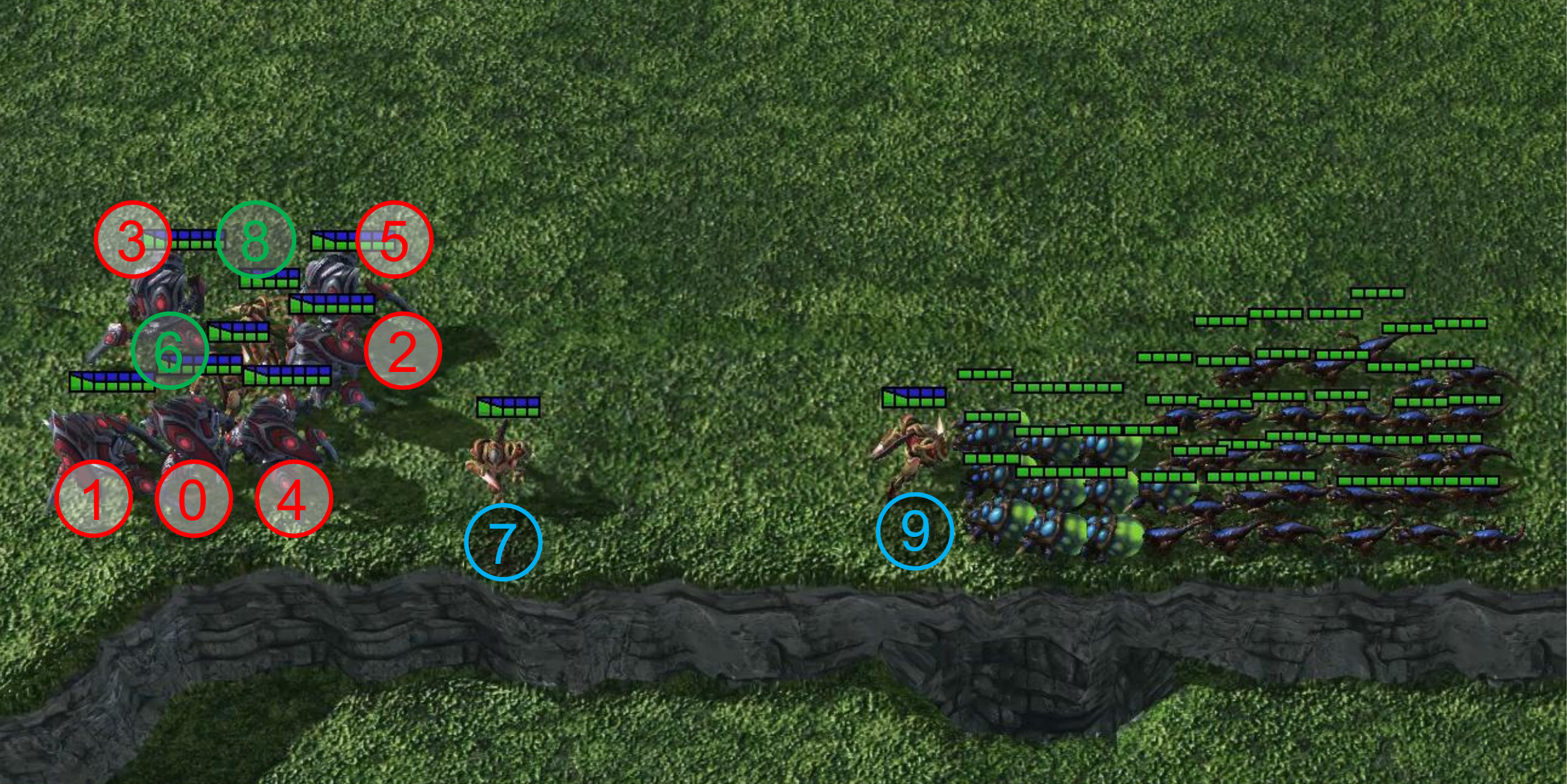}\hfill
\includegraphics[width=0.32\linewidth]{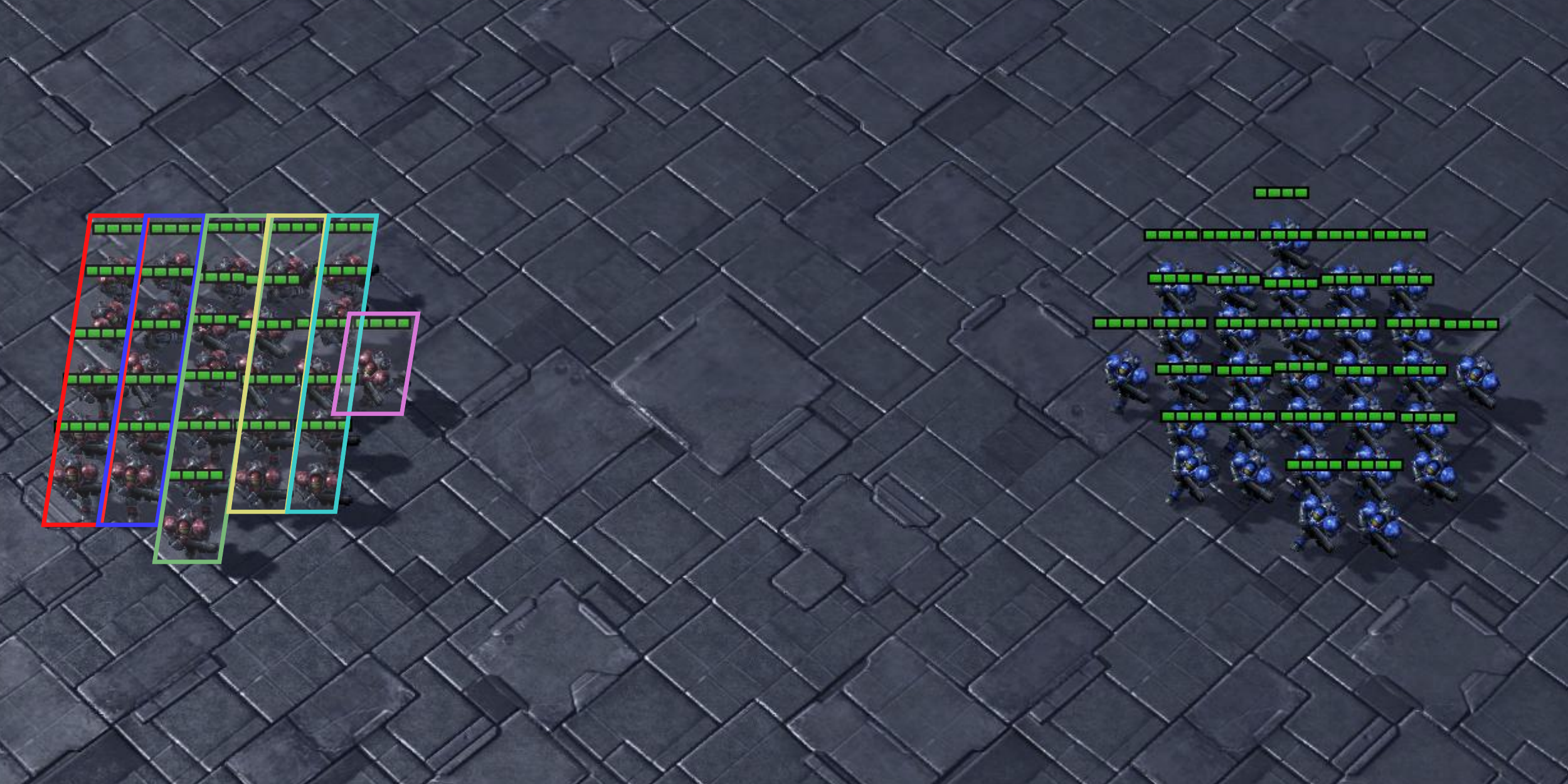}\hfill
\includegraphics[width=0.32\linewidth]{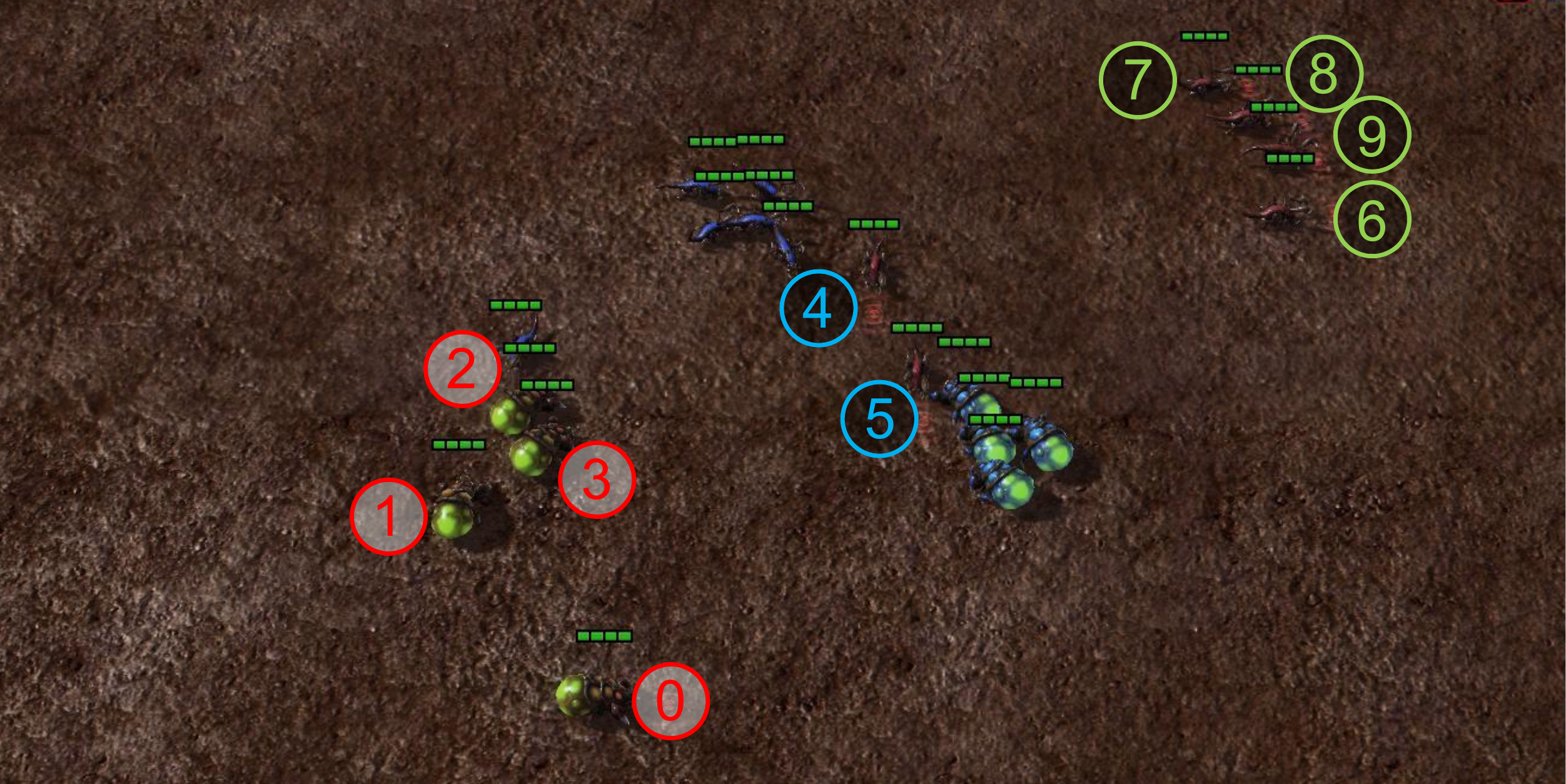}
\subfigure[Strategy: sacrificing Zealots 9 and 7 to minimize Banelings' splash damage. ]{\includegraphics[width=0.32\linewidth]{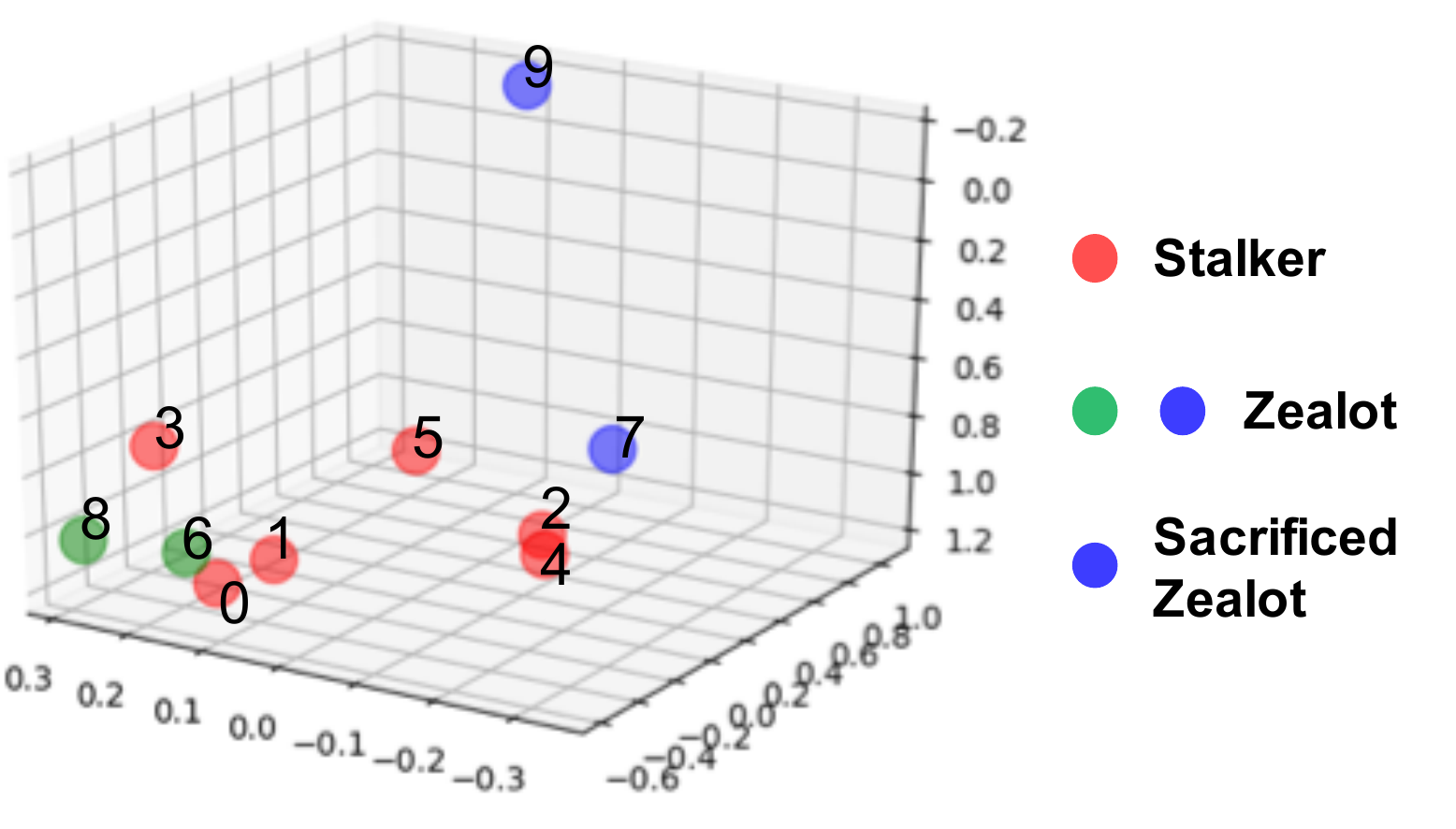}\label{fig:various_roles-6s4z_vs_10b30z}}\hfill
\subfigure[Strategy: forming an offensive concave arc quickly ]{\includegraphics[width=0.32\linewidth]{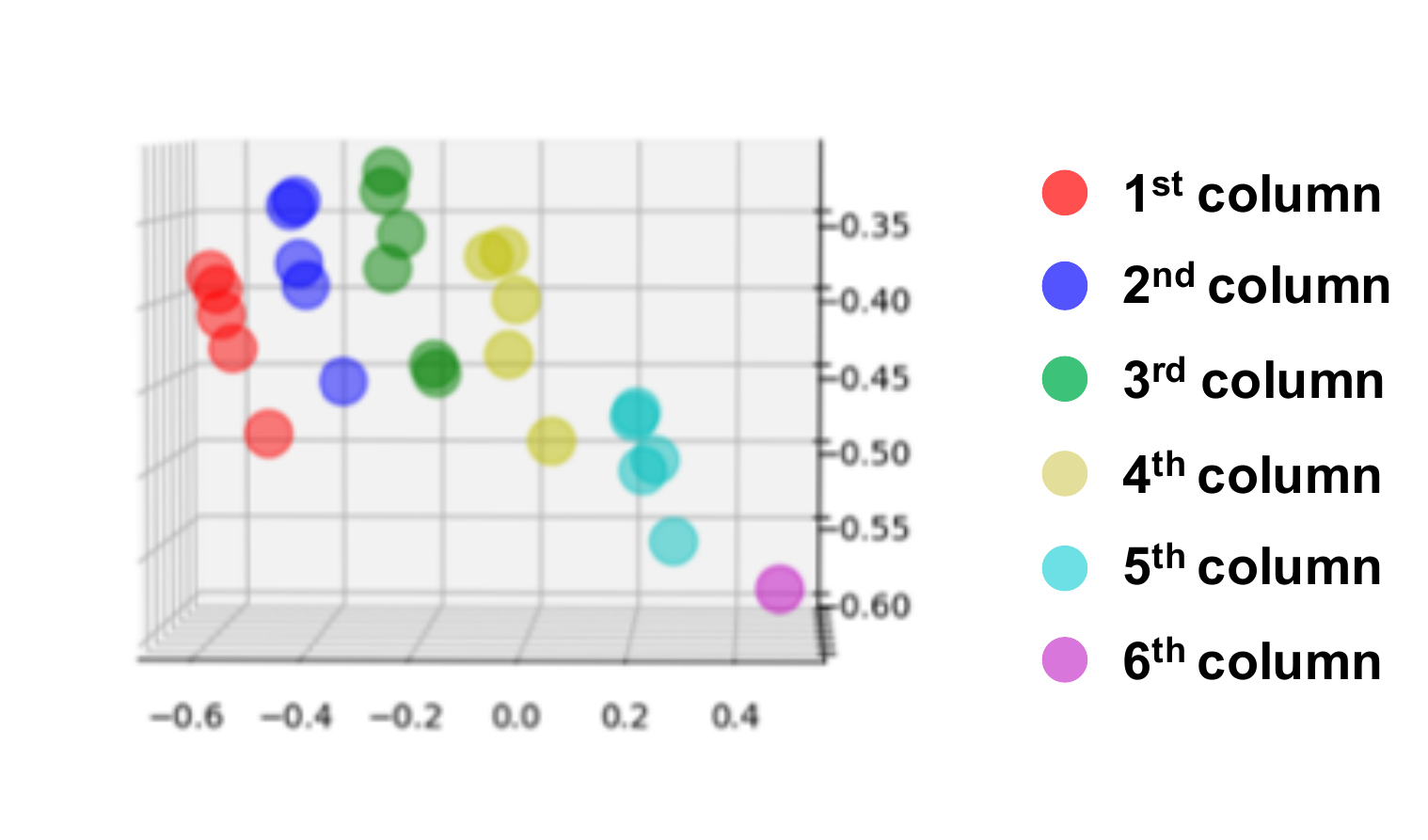}}\hfill
\subfigure[Strategy: green Zerglings hide away and Banelings kill most enemies by explosion.]{\includegraphics[width=0.32\linewidth]{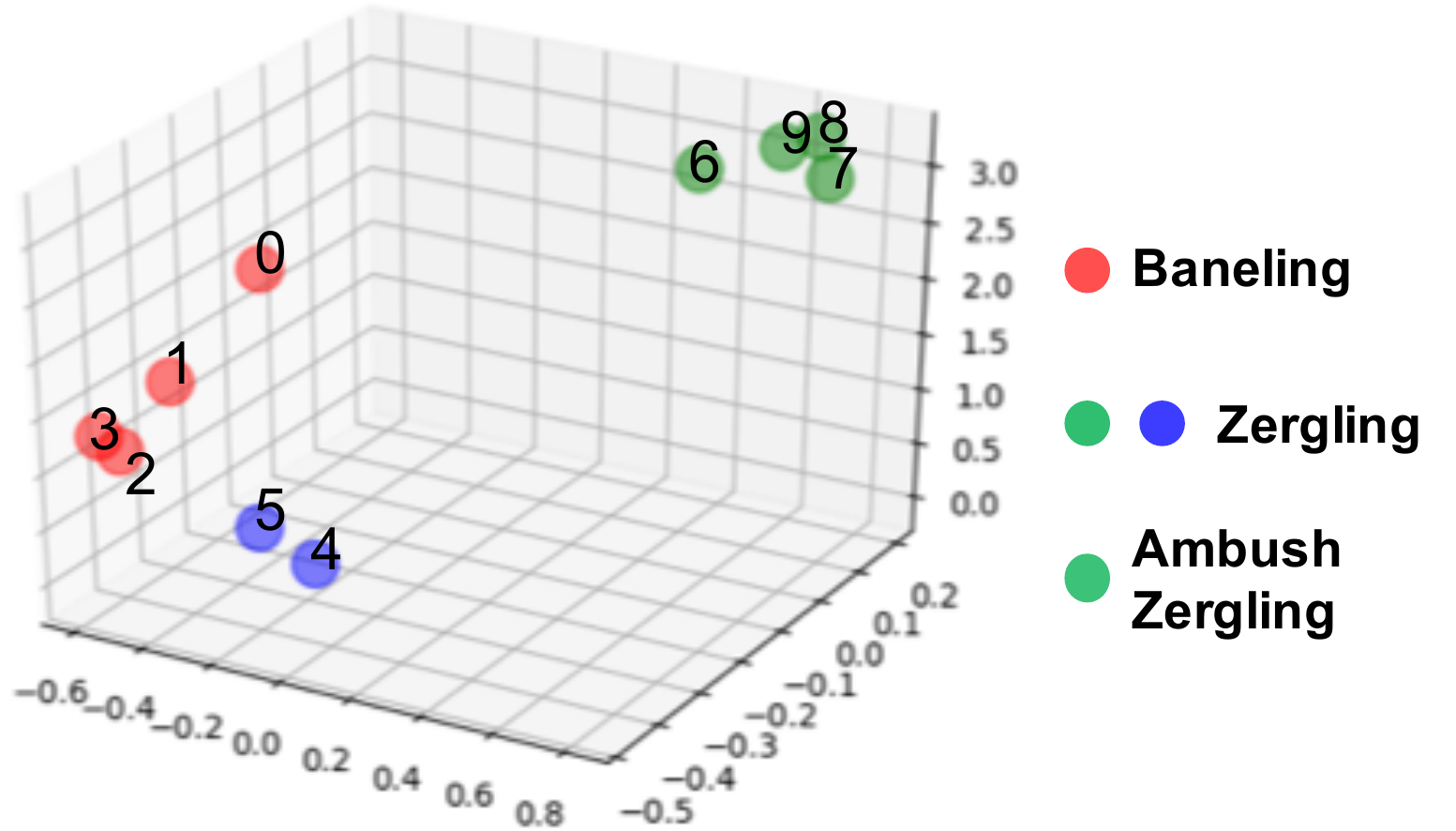}}
\caption{Learned roles for $\mathtt{6s4z\_vs\_10b30z}$, $\mathtt{27m\_vs\_30m}$, and $\mathtt{6z4b}$ (means of the role distributions, $\bm{\mu}_{\rho_i}$, are shown, without using any dimensionality reduction techniques), and the related, automatically discovered responsibilities.}\label{fig:various_roles}
\end{figure*}
\subsection{Dynamic Roles}\label{sec:exp-dynamic_role}
Answering the first and second questions, we show snapshots in an episode played by \name~agents on the StarCraft II micromanagement benchmark (SMAC) map $\mathtt{10m\_vs\_11m}$, where 10 Marines face 11 enemy Marines. As shown in Fig.~\ref{fig:dynamic_role-10m_vs_11m} (the role representations at $t$=$27$ are presented in Fig.~\ref{fig:teaser}), although observations contain much information, such as positions, health points, shield points, states of ally and enemy units, etc., the role encoder learns to focus on different parts of the observations according to the dynamically changed situations. At the beginning ($t$=$1$), agents need to form a concave arc to maximize the number of agents whose shoot range covers the front line of enemies. \name~learns to allocate roles according to agents' relative positions so that agents can quickly form the offensive formation using specialized policies. In the middle of the battle, one important tactic is to protect the injured ranged units. Our method learns this maneuver and roles cluster according to the remaining health points ($t$=$8$, $19$, $27$). Healthiest agents have role representations far from those of other agents. Such representations result in differentiated strategies: healthiest agents move forward to take on more firepower while other agents move backward, firing from a distance. In the meantime, some roles also cluster according to positions (agents 3 and 8 when $t$=$19$). The corresponding behaviors are agents with different roles fire alternatively to share the firepower. We can also observe that the role representations of dead agents aggregate together, representing a special group with an increasing number of agents during the battle.

These results demonstrate that our method learns dynamic roles and roles cluster clearly corresponding to automatically detected sub-tasks, in line with implicit constraints of the proposed optimization objectives.

\subsection{Performance on StarCraft II}\label{sec:exp-performance}
To test whether these roles and the corresponding sub-task specialization can improve learning efficiency, we test our method on the StarCraft II micromanagement (SMAC) benchmark~\cite{samvelyan2019starcraft}. This benchmark consists of various maps which have been classified as \emph{easy}, \emph{hard}, and \emph{super hard}. We compare \name~with algorithms shown in Table~\ref{tab:baselines} and present results for one \textbf{easy} map ($\mathtt{2s3z}$), three \textbf{hard} maps ($\mathtt{5m\_vs\_6m}$, $\mathtt{8m\_vs\_9m}$ \& $\mathtt{10m\_vs\_11m}$), and two \textbf{super hard} maps ($\mathtt{MMM2}$ \& $\mathtt{27m\_vs\_30m}$). Although SMAC benchmark is challenging, it is not specially designed to test performance in tasks with many agents. We thus introduce three new SMAC maps to test the scalability of our method, which are described in detail in Appendix C.
\begin{figure*}
    \centering
    \includegraphics[width=\linewidth]{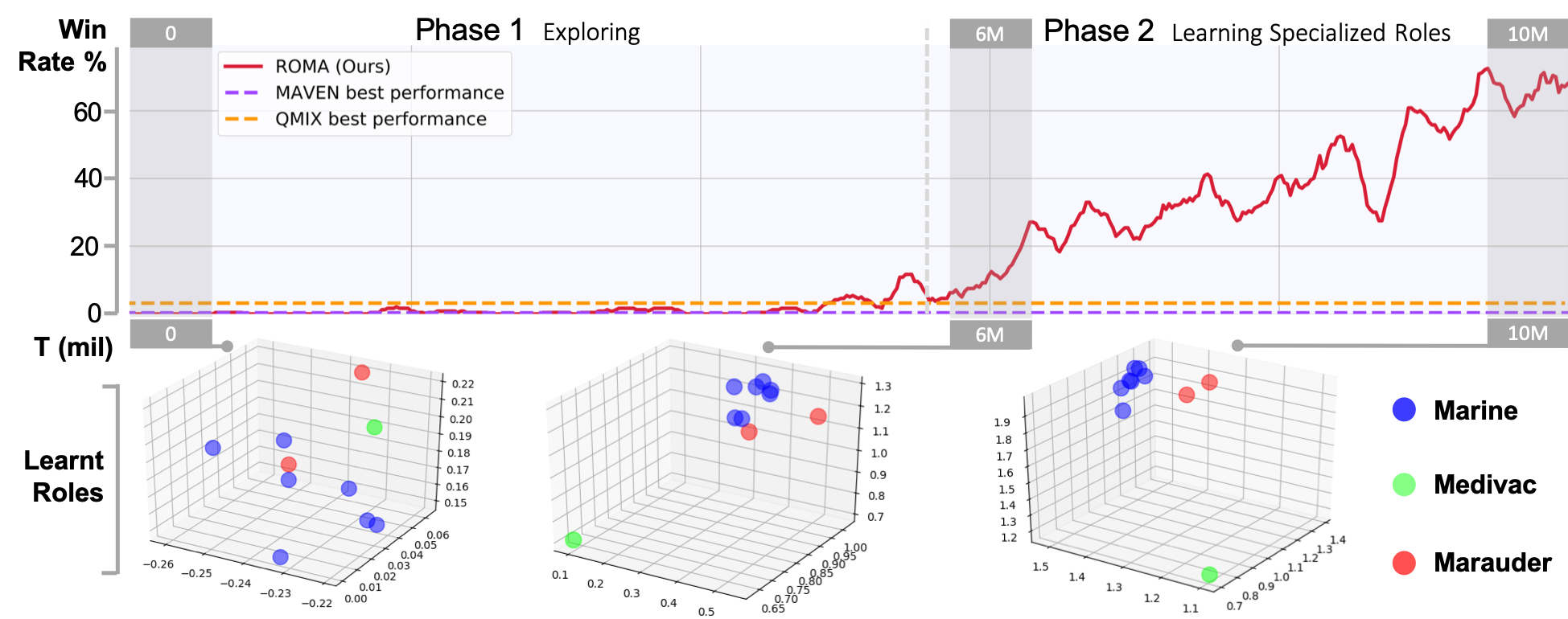}
    \caption{Role emergence and evolution on the map $\mathtt{MMM2}$ (role representations at time step $1$ are shown) during training (means of the role distributions, $\bm{\mu}_{\rho_i}$, are shown, without using any dimensionality reduction techniques). The emergence and specialization of roles is closely connected to the improvement of team performance. Agents in $\mathtt{MMM2}$ are heterogeneous, and we show role evolution process in a homogeneous team in Appendix C.3.}
    \label{fig:role_evolution}
\end{figure*}

For evaluation, all experiments in this section are carried out with 5 different random seeds, and results are shown with a $95\%$ confidence interval. Among these maps, four maps, $\mathtt{MMM2}$, $\mathtt{6s4z\_vs\_10b30z}$, $\mathtt{6z4b}$, and $\mathtt{10z5b\_vs\_2z3s}$, feature heterogeneous agents, and the others have homogeneous agents. Fig.~\ref{fig:performance-baselines} shows that our method yields substantially better results than all the alternative approaches on both homogeneous and heterogeneous maps (additional plots can be found in Appendix C.1). MAVEN overcomes the negative effects of QMIX's monotonicity constraint on exploration. However, it performs less satisfactorily than QMIX on most maps. We believe this is because agents start engaging in the battle immediately after spawning in SMAC maps, and exploration is not the critical factor affecting performance.

\textbf{Ablations}\ \ We carry out ablation studies, comparing with the ablations shown in Table~\ref{tab:baselines} and present results on three maps: $\mathtt{MMM2}$ (heterogeneous), $\mathtt{10z5b\_vs\_2s3z}$, and $\mathtt{10m\_vs\_11m}$ (homogeneous) in Fig.~\ref{fig:performance-ablations-losses} and~\ref{fig:performance-ablations}. The superiority of our method against $\mathcal{L}_{TD}$ highlights the contribution of the proposed regularizers -- $\mathcal{L}_{TD}$ performs even worse than QMIX on two of the three maps. By comparing ROMA with $\mathcal{L}_{TD}+\mathcal{L}_{I}$ and $\mathcal{L}_{TD}+\mathcal{L}_{D}$, we can conclude that the specialization loss $\mathcal{L}_{D}$ is more important in terms of performance improvements. Introducing $\mathcal{L}_I$ can make training more stable (for example, on the map $\mathtt{10m\_vs\_11m}$), but optimizing $\mathcal{L}_I$ alone can only slightly improve the performance. These observations support the claim that sub-task specialization can improve labor efficiency.

Comparison between QMIX-NPS and QMIX demonstrates that parameter sharing can, as documented~\cite{foerster2018counterfactual, rashid2018qmix}, speed up training. As discussed in the introduction, both these two paradigms may not get the best possible performance. In contrast, our method provides a dynamic learning sharing mechanism -- agents committed to a certain responsibility have similar policies. The comparison of the performance of \name, QMIX, and QMIX-NPS proves that such sub-task specialization can indeed improve team performance. What's more, comparison of~\name~against QMIX-LAR proves that the superiority of our method does not depend on the larger number of parameters. 

The performance gap between \name~and ablations is more significant on maps with more than ten agents. This observation supports discussions in previous sections -- the emergence of role is more likely to improve the labor efficiency in larger populations.

\subsection{Role Embedding Representations}\label{sec:exp-various_roles}
To explain the superiority of \name, we present the learned role embedding representations for three maps in Fig.~\ref{fig:various_roles}. Roles are representative of automatically discovered sub-tasks in the learned winning strategy. In the map of $\mathtt{6s4z\_vs\_10b30z}$, \name~learns to sacrifice Zealots 9 and 7 to kill all the enemy Banelings. Specifically, Zealots 9 and 7 will move to the frontier one by one to minimize the splash damage, while other agents will stay away and wait until all Banelings explode. Fig.~\ref{fig:various_roles-6s4z_vs_10b30z} shows the role embedding representations while performing the first sub-task where agent 9 is sacrificed. We can see that the role of Zealot 9 is quite different from those of other agents. Correspondingly, the strategy at this time is agent 9 moving rightward while other agents keep still. Detailed analysis for the other two maps can be found in Appendix C.2.

\subsection{Emergence and Evolution of Roles}\label{sec:exp-role_evolution}
We have shown the learned role representations and performance of our method, but the relationship between roles and performance remains unclear. To make up for this shortcoming, we visualize the emergence and evolution of roles during the training process on the map $\mathtt{MMM2}$ (heterogeneous) and $\mathtt{10m\_vs\_11m}$ (homogeneous). We discuss the results on $\mathtt{MMM2}$ here and defer analysis of $\mathtt{10m\_vs\_11m}$ to Appendix C.3.

In $\mathtt{MMM2}$, 1 Medivac, 2 Marauders, and 7 Marines are faced with a stronger enemy team consisting of 1 Medivac, 3 Marauders, and 8 Marines. Among the three involved unit types, Medivac is the most special one for that it can heal the injured units. In Fig.~\ref{fig:role_evolution}, we show one of the learning curves of \name~(red) and the role representations at the first environment step at three different stages. When the training begins ($T$=$0$), roles are random, and the agents are exploring the environment to learn the basic dynamics and the structure of the task. By $T$=$6$M, \name~has learned that the responsibilities of the Medivac are different from those of Marines and Marauders. The role, and correspondingly, the policy of the Medivac becomes quite different (Fig.~\ref{fig:role_evolution} middle). Such differentiation in behaviors enables agents to start winning the game. Gradually, \name~learns that Marines and Marauders have dissimilar characteristics and should take different sub-tasks, indicated by the differentiation of their role representations (Fig.~\ref{fig:role_evolution} right). This further specialization facilitates the performance increase between $6$M and $10$M. After $T$=$10$M, the responsibilities of roles are clear, and, as a result, the win rate gradually converges (Fig.~\ref{fig:performance-baselines} top left). For comparison, \name~without $\mathcal{L}_I$ and $\mathcal{L}_D$ can not even win once on this challenging task ($\mathcal{L}_{TD}$ in Fig.~\ref{fig:performance-ablations}-left). These results demonstrate that the gradually specialized roles are indispensable in team performance improvement.

\begin{table} [t]
    \caption{The mean and standard deviation of the learned dissimilarities $d_{\phi}$ between agents' trajectories on the map $\mathtt{MMM2}$.}
    \label{tab:specialized_roles}
    \centering
    \begin{tabular}{cr}
        \toprule
         Between different unit types& $0.9556\pm 0.0009$ \\
         Between the same unit type & $0.0780\pm 0.0019$ \\
        \toprule
    \end{tabular}
\end{table}
Moreover, we find that the learned dissimilarity model $d_{\phi}$ introduced in Sec.~\ref{sec:specialized_roles} provides an empirical evaluation for identifying new roles. We use the map $\mathtt{MMM2}$ as an example, where, as we discussed above, the learned roles of agents are characterized by their unit types. After scaling to $[0, 1]$, the learned dissimilarity between trajectories of agents with different unit types is close to $0.96$, while the learned dissimilarity between trajectories of agents with the same unit type is around $0.08$. These results indicate that an appropriate threshold can be used to decide when an individual behavior (trajectory) can be assigned the terminology \emph{role}.

In summary, our experiments demonstrate that \name~can learn dynamic, identifiable, versatile, and specialized roles that effectively decompose the task. Drawing support from these emergent roles, our method significantly pushes forward the state of the art of multi-agent reinforcement learning algorithms.

%% file: 6-ClosingRemarks.tex
\section{Closing Remarks}

We have introduced the concept of roles into deep multi-agent reinforcement learning by capturing the emergent roles and encouraging them to specialize on a set of automatically detected sub-tasks. Such deep role-oriented multi-agent learning framework provides another perspective to explain and promote cooperation within agent teams, and implicitly draws connection to the division of labor, which has been practiced in many natural systems for long.

To our best knowledge, this paper is making a first attempt at learning roles via deep reinforcement learning. The gargantuan task of understanding the emergence of roles, the division of labor, and interactions between more complex roles in hierarchical organization still lies ahead. We believe that these topics are basic and indispensable in building effective, flexible, and general-purpose multi-agent systems and this paper can help tackle these challenges.

%% file: A1-Math.tex
\section{Mathematical Derivation}

\subsection{Identifiable Roles}\label{sec:iv_roles}
For learning identifiable roles, we propose to maximize the conditional mutual information objective between roles and local observation-action histories given the current observations. In Sec. 3.1 of the paper, we introduce a posterior estimator and derive a tractable lower bound of the mutual information term:
\begin{align*}
   I(&\rho^t_i; \tau^{t-1}_i | o^t_i) = \mathbb{E}_{\rho^t_i, \tau^{t-1}_i, o^t_i}\left[\log\frac{p(\rho^t_i | \tau^{t-1}_i, o^t_i)}{p(\rho^t_i | o^t_i)}\right] \\
   & = \mathbb{E}_{\rho^t_i, \tau^{t-1}_i, o^t_i}\left[\log\frac{q_\xi(\rho^t_i | \tau^{t-1}_i, o^t_i)}{p(\rho^t_i | o^t_i)}\right] \\ & \ \ \ \ + \mathbb{E}_{\tau^{t-1}_i, o^t_i}\left[\KL(p(\rho^t_i|\tau^{t-1}_i, o^t_i) \| q_\xi(\rho^t_i|\tau^{t-1}_i, o^t_i))\right] \stepcounter{equation}\tag{\theequation} \\
   & \ge  \mathbb{E}_{\rho^t_i, \tau^{t-1}_i, o^t_i}\left[\log\frac{q_\xi(\rho^t_i | \tau^{t-1}_i, o^t_i)}{p(\rho^t_i | o^t_i)}\right],
   \label{aequ:mi}
\end{align*}
where the last inequality holds because of the non-negativity of the KL divergence. Then it follows that:
\begin{equation}
\begin{aligned}
   &\mathbb{E}_{\rho^t_i, \tau^{t-1}_i, o^t_i}\left[\log\frac{q_\xi(\rho^t_i | \tau^{t-1}_i, o^t_i)}{p(\rho^t_i | o^t_i)}\right]\\
  =&\mathbb{E}_{\rho^t_i, \tau^{t-1}_i, o^t_i}\left[\log q_\xi(\rho^t_i | \tau^{t-1}_i, o^t_i)\right]-\mathbb{E}_{\rho^t_i, o^t_i}\left[\log p(\rho^t_i | o^t_i)\right]\\
  =&\mathbb{E}_{\rho^t_i, \tau^{t-1}_i, o^t_i}\left[\log q_\xi(\rho^t_i | \tau^{t-1}_i, o^t_i)\right]+\mathbb{E}_{o^t_i}\left[H(\rho_i^t | o_i^t)\right] \\
  =&\mathbb{E}_{\tau^{t-1}_i, o^t_i}\left[\int p(\rho^t_i | \tau^{t-1}_i, o^t_i) \log q_\xi(\rho^t_i | \tau^{t-1}_i, o^t_i)d\rho^t_i\right] +\mathbb{E}_{o^t_i}\left[H(\rho_i^t | o_i^t)\right]
   \label{aequ:mi_derivation}
\end{aligned}
\end{equation}
The role encoder is conditioned on the local observations, so given the observations, the distributions of roles, $p(\rho^t_i)$, are independent from the local histories. Thus, we have
\begin{equation}
\begin{aligned}
      I(\rho^t_i; \tau^{t-1}_i | o^t_i) 
  \ge -\mathbb{E}_{\tau^{t-1}_i, o^t_i}\left[\mathcal{CE}[p(\rho^t_i | o^t_i) \| q_\xi(\rho^t_i | \tau^{t-1}_i, o^t_i)\right] +    \mathbb{E}_{o^t_i}\left[H(\rho_i^t | o_i^t)\right]
   \label{aequ:mi_assumption}
\end{aligned}
\end{equation}
In practice, we use a replay buffer $\mathcal{D}$ and minimize
\begin{equation}
\begin{aligned}
    \mathcal{L}_I(\theta_\rho, \xi) = \mathbb{E}_{(\tau^{t-1}_i, o^t_i)\sim\mathcal{D}}\left[\mathcal{CE}[p(\rho^t_i | o^t_i) \| q_\xi(\rho^t_i | \tau^{t-1}_i, o^t_i) - H(\rho_i^t | o_i^t)\right].
\end{aligned}
\end{equation}

\subsection{Specialized Roles}
Conditioning roles on local observations enables roles to be dynamic, and optimizing $\mathcal{L}_I$ enables roles to be identifiable by agents' long-term behaviors, but these formulations do not explicitly encourage specialized roles. To make up for this shortcoming, we propose a role differentiation objective in Sec. 3.2 of the paper, where a mutual information maximization objective is involved (maximizing $I(\rho^{t}_i; \tau^{t-1}_j | o^t_j)$). Here, we derive a variational lower bound of this mutual information objective to render it feasible to be optimized.
\begin{equation}
\begin{aligned}
I(\rho^{t}_i; \tau^{t-1}_j | o^t_j) &= \mathbb{E}_{\rho^{t}_i, \tau^{t-1}_j, o^t_j}\left[\log\frac{p(\rho^{t}_i, \tau^{t-1}_j | o^t_j)}{p(\rho^{t}_i | o^t_j)p(\tau^{t-1}_j | o^t_j)}\right] \\
&= \mathbb{E}_{\rho^{t}_i, \tau^{t-1}_j, o^t_j}\left[\log\frac{p(\rho^{t}_i| \tau^{t-1}_j, o^t_j)}{p(\rho^{t}_i | o^t_j)}\right] \\
&= \mathbb{E}_{\rho^{t}_i, \tau^{t-1}_j, o^t_j}\left[\log p(\rho^{t}_i| \tau^{t-1}_j, o^t_j)\right] + \mathbb{E}_{o^t_j}\left[H(\rho^{t}_i | o^t_j)\right]\\
&\ge \mathbb{E}_{\rho^{t}_i, \tau^{t-1}_j, o^t_j}\left[\log p(\rho^{t}_i| \tau^{t-1}_j, o^t_j)\right].
\end{aligned}
\end{equation}
\begin{figure*}
\includegraphics[width=\linewidth]{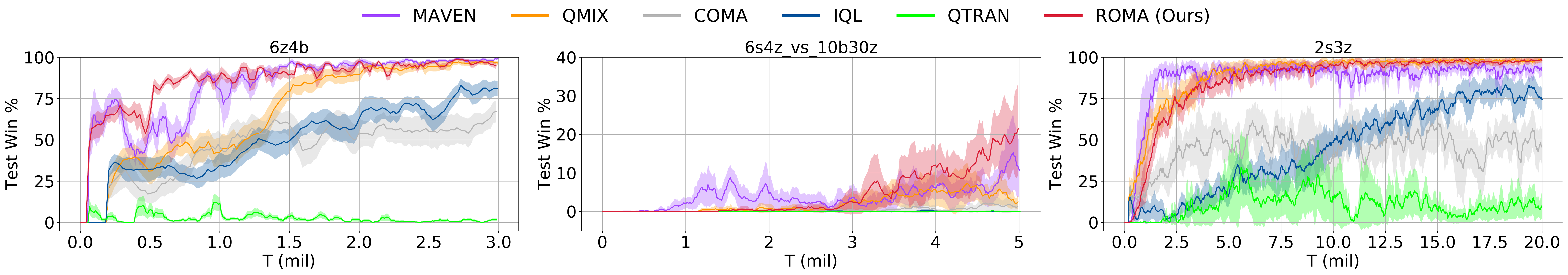}
\caption{Additional results on the SMAC benchmark.}\label{afig:additional_performance}
\end{figure*}

We clip the variances of role distributions at a small value ($0.1$) to ensure that the entropy of role distributions are always non-negative so that the last inequality holds. Then, it follows that:
\begin{equation}
\begin{aligned}
&\mathbb{E}_{\rho^{t}_i, \tau^{t-1}_j, o^t_j}\left[\log p(\rho^{t}_i| \tau^{t-1}_j, o^t_j)\right] \\
=& \mathbb{E}_{\rho^{t}_i, \tau^{t-1}_j, o^t_j}\left[\log q_\xi(\rho^{t}_i| \tau^{t-1}_j, o^t_j)\right] + \mathbb{E}_{\tau^{t-1}_j, o^t_j}\left[\KL\left[p(\rho^{t}_i| \tau^{t-1}_j, o^t_j) \| q_\xi(\rho^{t}_i| \tau^{t-1}_j, o^t_j) \right]\right] \\
\ge & \mathbb{E}_{\rho^{t}_i, \tau^{t-1}_j, o^t_j}\left[\log q_\xi(\rho^{t}_i| \tau^{t-1}_j, o^t_j)\right]\label{aequ:s_lower_bound},
\end{aligned}
\end{equation}
where $q_\xi$ is the trajectory encoder introduced in Sec.~\ref{sec:iv_roles}, and the KL divergence term can be left out when deriving the lower bound because it is non-negative. Therefore, we have:
\begin{equation}
\begin{aligned}
I(\rho^{t}_i&; \tau^{t-1}_j | o^t_j) \ge \mathbb{E}_{\rho^{t}_i, \tau^{t-1}_j, o^t_j}\left[\log q_\xi(\rho^{t}_i| \tau^{t-1}_j, o^t_j)\right].\label{aequ:t1get2}
\end{aligned}
\end{equation}
Recall that, in order to learn specialized roles, we propose to minimize:
\begin{equation}
\begin{aligned}
\|D_\phi^t\|_{F} - \sum_{i\ne j}\min\{I(\rho^{t}_i; \tau^{t-1}_j|o^t_j) + d_\phi(\tau^{t-1}_i, \tau^{t-1}_j), U\},\label{aequ:raw_obj}
\end{aligned}
\end{equation}
where $D^t_\phi=(d^t_{ij})$, and $d^t_{ij}=d_\phi(\tau^{t-1}_i, \tau^{t-1}_j)$ is the estimated dissimilarity between trajectories of agent $i$ and $j$. For the term $\min\{I(\rho^{t}_i; \tau^{t-1}_j|o^t_j) + d_\phi(\tau^{t-1}_i, \tau^{t-1}_j), U\}$, we have:
\begin{equation}
\begin{aligned}
&\min\{I(\rho^{t}_i; \tau^{t-1}_j|o^t_j) + d_\phi(\tau^{t-1}_i, \tau^{t-1}_j), U\} \\ = &\min\{\mathbb{E}_{\bm{\tau}^{t-1}, \bm{o}^t, \bm{\rho}^t}\left[\log\frac{p(\rho^{t}_i, \tau^{t-1}_j | o^t_j)}{p(\rho^{t}_i | o^t_j)p(\tau^{t-1}_j | o^t_j)} + d_\phi(\tau^{t-1}_i, \tau^{t-1}_j)\right], \mathbb{E}_{\bm{\tau}^{t-1}, \bm{o}^t, \bm{\rho}^t}\left[U\right]\},
\end{aligned}
\end{equation}
where $\bm{\tau}^{t-1}$ is the joint trajectory, $\bm{o}^{t}$ is the joint observation, and $\bm{\rho}^{t}=\langle \rho_1^t, \rho_2^t, \cdots, \rho_n^t\rangle$. We denote 
\begin{equation}
\begin{aligned}
T_1 &\equiv \log\frac{p(\rho^{t}_i, \tau^{t-1}_j | o^t_j)}{p(\rho^{t}_i | o^t_j)p(\tau^{t-1}_j | o^t_j)},\\
T_2 &\equiv \log q_\xi(\rho^{t}_i| \tau^{t-1}_j, o^t_j).
\end{aligned}
\end{equation}
Because
\begin{equation}
\begin{aligned}
T_2 &\ge \min\{T_2, U\}, \\
U &\ge \min\{T_2, U\},
\end{aligned}
\end{equation}
it follows that:
\begin{equation}
\begin{aligned}
\mathbb{E}_{\bm{\tau}^{t-1}, \bm{o}^t, \bm{\rho}^t}\left[T_2\right] &\ge \mathbb{E}_{\bm{\tau}^{t-1}, \bm{o}^t, \bm{\rho}^t}\left[\min\{T_2, U\}\right], \\
\mathbb{E}_{\bm{\tau}^{t-1}, \bm{o}^t, \bm{\rho}^t}\left[U\right] &\ge \mathbb{E}_{\bm{\tau}^{t-1}, \bm{o}^t, \bm{\rho}^t}\left[\min\{T_2, U\}\right].\label{aequ:xygemin}
\end{aligned}
\end{equation}
So that
\begin{equation}
\begin{aligned}
 &\min\{\mathbb{E}_{\bm{\tau}^{t-1}, \bm{o}^t, \bm{\rho}^t}\left[T_1\right], \mathbb{E}_{\bm{\tau}^{t-1}, \bm{o}^t, \bm{\rho}^t}\left[U\right]\} \\
 \ge &\min\{\mathbb{E}_{\bm{\tau}^{t-1}, \bm{o}^t, \bm{\rho}^t}\left[T_2\right], \mathbb{E}_{\bm{\tau}^{t-1}, \bm{o}^t, \bm{\rho}^t}\left[U\right]\} \ \ \ \ \{\text{Eq.~\ref{aequ:t1get2}}\} \\
 \ge & \mathbb{E}_{\bm{\tau}^{t-1}, \bm{o}^t, \bm{\rho}^t}\left[\min\{T_2, U\}\right] \ \ \ \ \{\text{Eq.~\ref{aequ:xygemin}}\}, \label{aequ:minexpectation}
\end{aligned}
\end{equation}
which means that Eq.~\ref{aequ:raw_obj} satisfies:
\begin{align*}
  &\|D_\phi^t\|_{F} - \sum_{i\ne j}\min\{I(\rho^{t}_i; \tau^{t-1}_j|o^t_j) + d_\phi(\tau^{t-1}_i, \tau^{t-1}_j), U\} \\
= & \mathbb{E}_{\bm{\tau}^{t-1}, \bm{o}^t, \bm{\rho}^t}\left[\|D_\phi^t\|_{F}\right] - \sum_{i\ne j}\min \{ \mathbb{E}_{\bm{\tau}^{t-1}, \bm{o}^t, \bm{\rho}^t}\left[T_1+d_\phi(\tau^{t-1}_i, \tau^{t-1}_j)\right], U\}\\ \stepcounter{equation}\tag{\theequation}\label{aequ:specialized_math_lower_bound}
\le & \mathbb{E}_{\bm{\tau}^{t-1}, \bm{o}^t, \bm{\rho}^t}\left[\|D_\phi^t\|_{F}\right] - \sum_{i\ne j}\mathbb{E}_{\bm{\tau}^{t-1}, \bm{o}^t, \bm{\rho}^t}\left[\min \{ T_2+d_\phi(\tau^{t-1}_i, \tau^{t-1}_j), U\}\right] \ \ \ \ \{\text{Eq.~\ref{aequ:minexpectation}}\}\\
= & \mathbb{E}_{\bm{\tau}^{t-1}, \bm{o}^t, \bm{\rho}^t}\left[\|D_\phi^t\|_{F} - \sum_{i\ne j}\min \{ T_2+d_\phi(\tau^{t-1}_i, \tau^{t-1}_j), U\}\right].
\end{align*}
We minimize this upper bound to optimize Eq.~\ref{aequ:raw_obj}. In practice, we use a replay buffer, and minimize:
\begin{equation}
\begin{aligned}
    \mathcal{L}_D(\theta_\rho, \phi, \xi) = \mathbb{E}_{(\bm{\tau}^{t-1}, \bm{o}^t)\sim\mathcal{D}, \bm{\rho}^t\sim p(\bm{\rho}^t | \bm{o}^t)}\big[\|D_\phi^t\|_{F} - \sum_{i\ne j}\min\{q_\xi(\rho^{t}_i| \tau^{t-1}_j, o^t_j) + d_\phi(\tau^{t-1}_i, \tau^{t-1}_j), U\}\big],
\end{aligned}
\end{equation}
where $\mathcal{D}$ is the replay buffer, $\bm{\tau}^{t-1}$ is the joint trajectory, $\bm{o}^{t}$ is the joint observation, and $\bm{\rho}^{t}=\langle \rho_1^t, \rho_2^t, \cdots, \rho_n^t\rangle$.

%% file: A2-Architecture.tex
\begin{figure*}
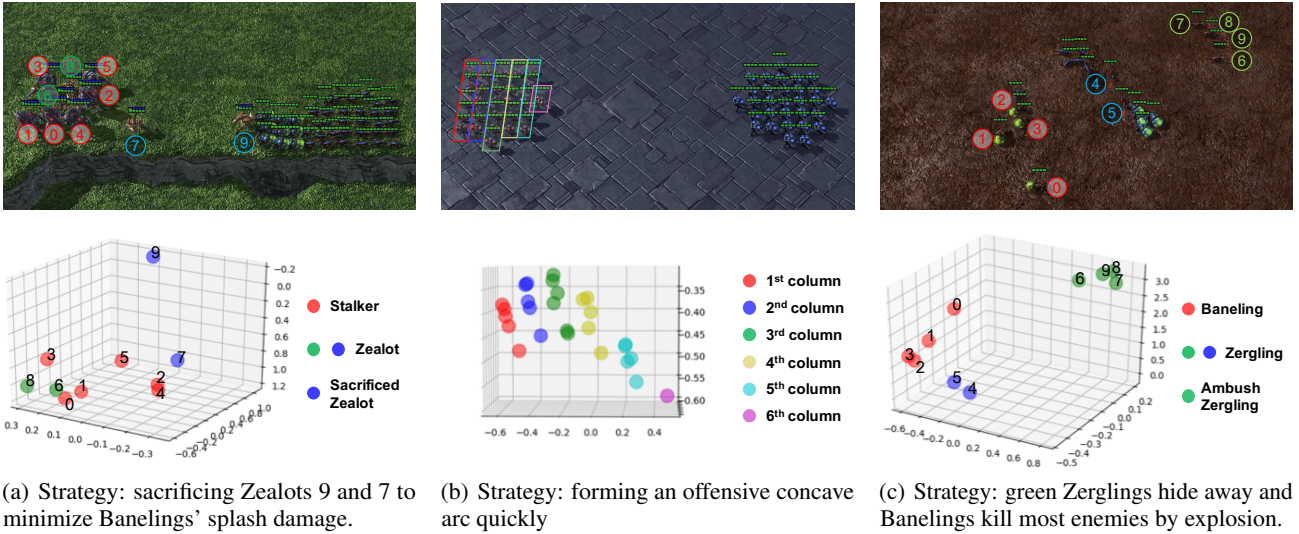

\centering
\includegraphics[width=0.32\linewidth]{fig-various_roles/various_roles-g1.pdf}\hfill
\includegraphics[width=0.32\linewidth]{fig-various_roles/various_roles-g2.pdf}\hfill
\includegraphics[width=0.32\linewidth]{fig-various_roles/various_roles-g3.pdf}
\subfigure[Strategy: sacrificing Zealots 9 and 7 to minimize Banelings' splash damage. ]{\includegraphics[width=0.32\linewidth]{fig-various_roles/various_roles-r1.pdf}\label{afig:various_roles-6s4z_vs_10b30z}}\hfill
\subfigure[Strategy: forming an offensive concave arc quickly ]{\includegraphics[width=0.32\linewidth]{fig-various_roles/various_roles-r2.pdf}\label{afig:various_roles-27m_vs_30m}}\hfill
\subfigure[Strategy: green Zerglings hide away and Banelings kill most enemies by explosion.]{\includegraphics[width=0.32\linewidth]{fig-various_roles/various_roles-r3.pdf}\label{afig:various_roles-6z4b}}
\caption{(Reproduced from Fig. 6 in the paper, for quick reference.) Learned roles for $\mathtt{6s4z\_vs\_10b30z}$, $\mathtt{27m\_vs\_30m}$, and $\mathtt{6z4b}$ (means of the role distributions, $\bm{\mu}_{\rho_i}$, are shown, without using any dimensionality reduction techniques), and the related, automatically discovered responsibilities.}\label{afig:various_roles}
\end{figure*}

\section{Architecture, Hyperparameters, and Infrastructure}

\subsection{\name}
In this paper, we base our algorithm on QMIX~\cite{rashid2018qmix}, whose framework is shown in Fig.~\ref{afig:qmix_framework} and described in Appendix~\ref{sec:appendix_related_works}. In \name, each agent has a neural network to approximate its local utility. The local utility network consists of three layers, a fully-connected layer, followed by a 64 bit GRU, and followed by another fully-connected layer that outputs an estimated value for each action. The local utilities are fed into a mixing network estimating the global action value. The mixing network has a 32-dimensional hidden layer with ReLU activation. Parameters of the mixing network are generated by a hyper-net conditioning on the global state. This hyper-net has a fully-connected hidden layer of 32 dimensions. These settings are the same as QMIX. 

We use very simple network structures for the components related to role embedding learning, i.e., the role encoder, the role decoder, and the trajectory encoder. The multi-variate Gaussian distributions from which the individual roles are drawn have their means and variances generated by the role encoder, which is a fully-connected network with a 12-dimensional hidden layer with ReLU activation. The parameters in the second fully-connected layers of the local utility approximators are generated by the role decoder whose inputs are the individual roles, which are 3-dimensional in all experiments. The role decoder is also a fully-connected network with a 12-dimensional hidden layer and ReLU activation. For the trajectory encoder, we again use a fully-connected network with a 12-dimensional hidden layer and ReLU activation. The inputs of the trajectory encoder are the hidden states of the GRUs in the local utility functions after the last time step.

For all experiments, we set $\lambda_I=10^{-4}$, $\lambda_D=10^{-2}$, and the discount factor $\gamma=0.99$. The optimization is conducted using RMSprop with a learning rate of $5\times 10^{-4}$, $\alpha$ of 0.99, and with no momentum or weight decay. For exploration, we use $\epsilon$-greedy with $\epsilon$ annealed linearly from $1.0$ to $0.05$ over $50k$ time steps and kept constant for the rest of the training. We run $8$ parallel environments to collect samples. Batches of 32 episodes are sampled from the replay buffer, and the whole framework is trained end-to-end on fully unrolled episodes. All experiments on StarCraft II use the default reward and observation settings of the SMAC benchmark.

Experiments are carried out on NVIDIA GTX 2080 Ti GPU.

\subsection{Baselines and Ablations}
We compare \name~with various baselines and ablations, which are listed in Table. 1 of the paper. For COMA~\cite{foerster2018counterfactual}, QMIX~\cite{rashid2018qmix}, and MAVEN~\cite{mahajan2019maven}, we use the codes provided by the authors where the hyper-parameters have been fin-tuned on the SMAC benchmark. QMIX-NPS uses the identical architecture as QMIX, and the only difference lies in that QMIX-NPS does not share parameters among agents. Compared to QMIX, for the local utility function of agents, QMIX-LAR adds two more fully-connected layers of 80 and 25 dimensions after the GRU layer so that it approximately has the same number of parameters as ROMA.

%% file: A3-Experiments.tex
\begin{figure*}
\includegraphics[width=\linewidth]{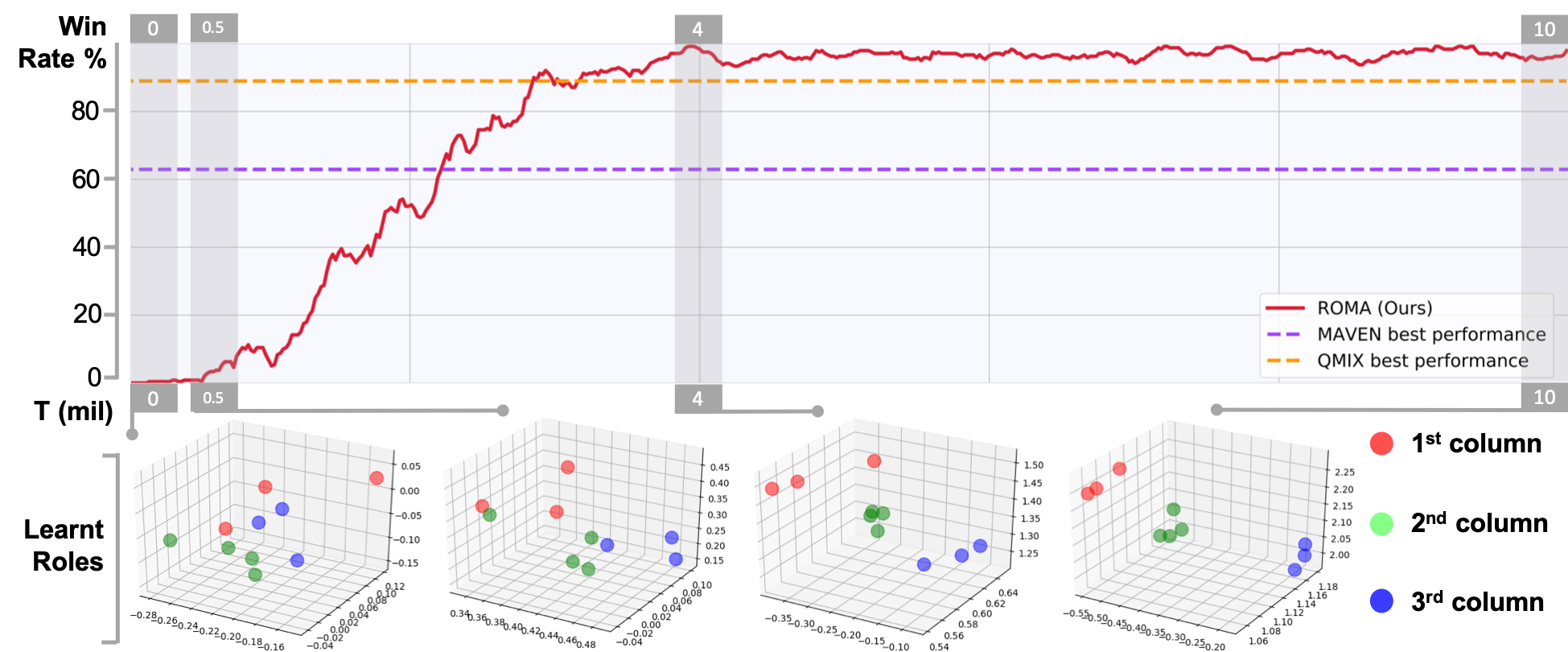}
\caption{The process of role emergence and evolution on the map $\mathtt{10m\_vs\_11m}$.}\label{afig:additional_evolution}
\end{figure*}

\section{Additional Experimental Results}

We benchmark our method on the StarCraft II unit micromanagement tasks. To test the scalability of the proposed approach, we introduce three maps. The $\mathtt{6z4b}$ map features symmetry teams consisting of 4 Banelings and 6 Zerglings. In the map of $\mathtt{6s4z\_vs\_10b30z}$, 6 Stalkers and 4 Zealots learn to defeat 10 Banelings and 30 Zerglings. And $\mathtt{10z5b\_vs\_2z3s}$ characterizes asymmetry teams consisting of 10 Zerglings \& 5 Banelings and 2 Zealots \& 3 Stalkers, respectively. 

\subsection{Performance Comparison against Baselines}
Fig.~\ref{afig:additional_performance} presents performance of \name~against various baselines on three maps. Performance comparison on the other maps is shown in Fig. 4 of the paper. We can see that the advantage of \name~is more significant on maps with more agents, such as $\mathtt{10z5b\_vs\_2z3s}$, $\mathtt{MMM2}$, $\mathtt{27\_vs\_30m}$, and $\mathtt{10m\_vs\_11m}$.

\subsection{Role Embedding Representations}
Fig.~\ref{fig:various_roles} shows various roles learned by \name. Roles are closely related to the sub-tasks in the learned winning strategy.

For the map $\mathtt{27m\_vs\_30m}$, the winning strategy is to form an offensive concave arc before engaging in the battle. Fig.~\ref{afig:various_roles-27m_vs_30m} illustrates the role embedding representations at the first time step when the agents are going to set up the attack formation. We can see the roles aggregate according to the relative positions of the agents. Such role differentiation leads to different moving strategies so that agents can quickly form the arc without collisions.

Similar role-behavior relationships can be seen in all tasks. We present another example on the task of $\mathtt{6z4b}$. In the winning strategy learned by \name, Zerglings 4 \& 5 and Banelings kill most of the enemies, taking advantage of the splash damage of the Banelings, while Zerglings 6-9 hideaway, wait until the explosion is over, and then kill the remaining enemies. Fig.~\ref{afig:various_roles-6z4b} shows the role embedding representations before the explosion. We can see clear clusters closely corresponding to the automatically detected sub-tasks at this time step.

Supported by these results, we can conclude that \name~can automatically decompose the task and learn versatile roles, each of which is specialized in a certain sub-task. 

\subsection{Additional Results for Role Evolution}
\begin{wrapfigure}{r}{0.38\linewidth}
    \centering
    \includegraphics[width=\linewidth]{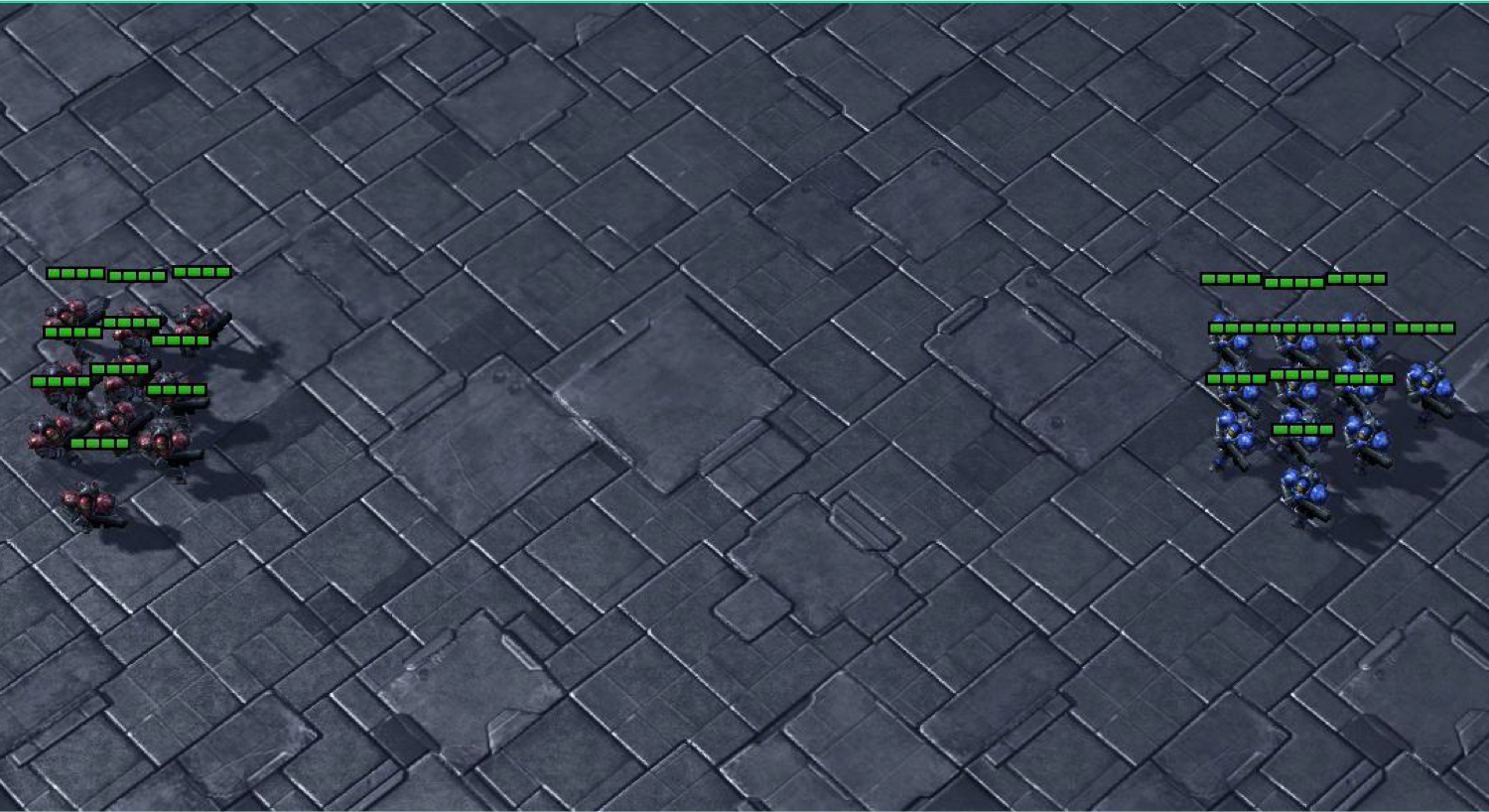}
    \caption{Screenshot of $\mathtt{10m\_vs\_11m}$, $t$=$1$.}
    \label{afig:10m_vs_11m-t1}
\end{wrapfigure}
In Fig. 7 of the paper, we show how roles emerge and evolve on the map $\mathtt{MMM2}$, where the involved agents are heterogeneous. In this section, we discuss the case of homogeneous agent teams. To this end, we visualize the emergence and evolution process of roles on the map $\mathtt{10m\_vs\_11m}$, which features 10 ally Marines facing 11 enemy Marines. In Fig.~\ref{afig:additional_evolution}, we show the roles at the first time step of the battle (screenshot can be found in Fig.~\ref{afig:10m_vs_11m-t1}) at four different stages during the training. At this moment, agents need to form an offensive concave arc quickly. We can see that \name~gradually learns to allocate roles according to relative positions of agents. Such roles and the corresponding differentiation in the individual policies help agents form the offensive arc more efficiently. Since setting up an attack formation is critical for winning the game, a connection between the specialization of the roles at the first time step and the improvement of the win rate can be observed.

%% file: A4-RelatedWorks.tex
\begin{figure*}
    \centering
    \includegraphics[width=0.8\linewidth]{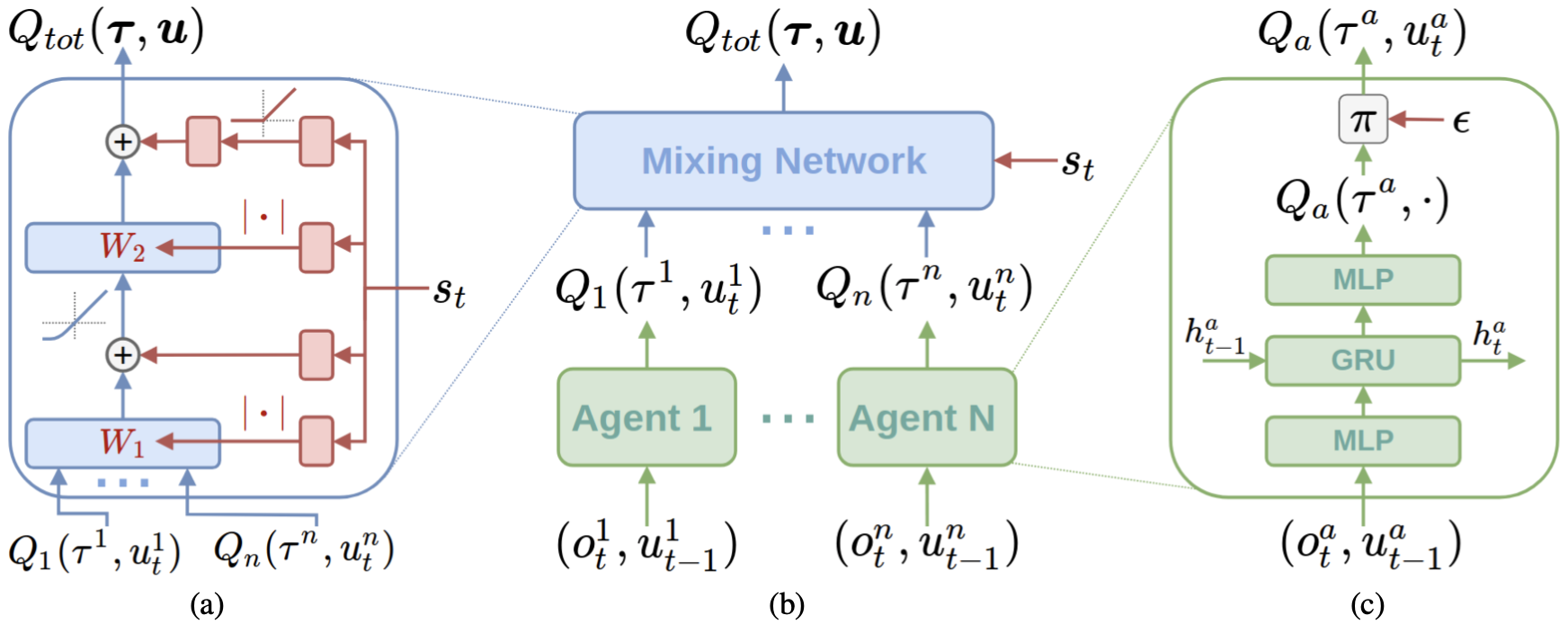}
    \caption{The framework of QMIX, reproduced from the original paper~\cite{rashid2018qmix}. (a) The architecture of the mixing network (blue), whose weights and biases are generated by a hyper-net (red) conditioned on the global state. (b) The overall QMIX structure. (c) Local utility network structure.}
    \label{afig:qmix_framework}
\end{figure*}

\section{Related Works}\label{sec:appendix_related_works}
Multi-agent reinforcement learning holds the promise to solve many real-world problems and has been making vigorous progress recently. To avoid otherwise exponentially large state-action space, factorizing MDPs for multi-agent systems is proposed~\cite{guestrin2002multiagent}. Coordination graphs~\cite{bargiacchi2018learning, yang2018glomo, grover2018evaluating, kipf2018neural} and explicit communication ~\cite{sukhbaatar2016learning, hoshen2017vain, jiang2018learning, singh2019learning, das2019tarmac, singh2019learning, kim2019learning} are studied to model the dependence between the decision-making processes of agents. Training decentralized policies is faced with two challenges: the issue of non-stationarity~\cite{tan1993multi} and reward assignment~\cite{foerster2018counterfactual, nguyen2018credit}. To resolve these problems, Sunehag et~al.~\yrcite{sunehag2018value} propose a value decomposition method called VDN. VDN learns a global action-value function, which is factored as the sum of each agent's local Q-value. QMIX~\cite{rashid2018qmix} extends VDN by representing the global value function as a learnable, state-condition, and monotonic combination of the local Q-values. In this paper, we use the mixing network of QMIX. The framework of QMIX is shown in Fig.~\ref{afig:qmix_framework}.

The StarCraft II unit micromanagement task is considered as one of the most challenging cooperative multi-agent testbeds for its high degree of control complexity and environmental stochasticity. Usunier et~al.~\yrcite{usunier2017episodic} and Peng et~al.~\yrcite{peng2017multiagent} study this problem from a centralized perspective. In order to facilitate decentralized control, we test our method on the SMAC benchmark~\cite{samvelyan2019starcraft}, which is the same as in \cite{foerster2017stabilising, foerster2018counterfactual, rashid2018qmix, mahajan2019maven}.